%% file: paper_new2.tex
\RequirePackage{lineno}
\documentclass[twocolumn,showpacs,aps,prd,floatfix,altaffilletter,altaffillsymbol]{revtex4}
\usepackage{graphicx}
\usepackage{dcolumn}
\usepackage{amsmath}
\usepackage{epsfig}
\usepackage{subfigure}
\usepackage{longtable}
\usepackage{bm}
\usepackage{relsize}
\input{babarsym}

\def\babar{\mbox{\slshape B\kern -0.1em{\smaller A}\kern-0.1emB\kern-0.1em{\smaller A\kern -0.2em R }}}

\def\en         {\ensuremath{e^-}\xspace}  
\def\ep         {\ensuremath{e^+}\xspace}

\def\ee         {\ensuremath{e^+e^-}\xspace}

\def\mup        {\ensuremath{\mu^+}\xspace}
\def\mun        {\ensuremath{\mu^-}\xspace} 
\def\mumu       {\ensuremath{\mu^+\mu^-}\xspace}

\def\taup       {\ensuremath{\tau^+}\xspace}
\def\taum       {\ensuremath{\tau^-}\xspace}

\def\ellell     {\ensuremath{\ell^+ \ell^-}\xspace}
\def\nub        {\ensuremath{\overline{\nu}}\xspace}
\def\nunub      {\ensuremath{\nu{\overline{\nu}}}\xspace}
\def\nub        {\ensuremath{\overline{\nu}}\xspace}
\def\nunub      {\ensuremath{\nu{\overline{\nu}}}\xspace}
\def\nue        {\ensuremath{\nu_e}\xspace}
\def\nueb       {\ensuremath{\nub_e}\xspace}

\def\num        {\ensuremath{\nu_\mu}\xspace}
\def\numb       {\ensuremath{\nub_\mu}\xspace}

\def\nut        {\ensuremath{\nu_\tau}\xspace}
\def\nutb       {\ensuremath{\nub_\tau}\xspace}

\def\nul        {\ensuremath{\nu_\ell}\xspace}
\def\nulb       {\ensuremath{\nub_\ell}\xspace}

\def\g     {\ensuremath{\gamma}\xspace}
\def\gaga  {\ensuremath{\gamma\gamma}\xspace}  
\def\ggstar{\ensuremath{\gamma\gamma^*}\xspace}

\def\Y#1S{\ensuremath{\Upsilon{(#1S)}}\xspace}

\def\FourS {\Y4S}

\def\mes        {\mbox{$m_{\rm ES}$}\xspace}
\def\DeltaE     {\mbox{$\Delta E$}\xspace}
\def\mev   {\mbox{MeV}\xspace}
\def\mevc   {\mbox{MeV/$c$}\xspace}
\def\mevcc   {\mbox{MeV/$c^{\rm 2}$}\xspace}

\def\gevcc   {\mbox{GeV/$c^{\rm 2}$}\xspace}
\def\Kp    {\ensuremath{K^+}\xspace}
\def\Km    {\ensuremath{K^-}\xspace}
\def\piz   {\ensuremath{\pi^0}\xspace}
\def\pip   {\ensuremath{\pi^+}\xspace}
\def\pim   {\ensuremath{\pi^-}\xspace}
\def\KS    {\ensuremath{K^0_{\scriptscriptstyle S}}\xspace}
\def\KL    {\ensuremath{K^0_{\scriptscriptstyle L}}\xspace}
\def\BR         {{\ensuremath{\cal B}\xspace}}
\def\jpsi     {\ensuremath{{J\mskip -3mu/\mskip -2mu\psi\mskip 2mu}}\xspace}

\def\Bz      {\ensuremath{B^0}\xspace}
\def\Bzb     {\ensuremath{\Bbar^0}\xspace}

\def\figurebox#1#2#3{
    \def\arg{#3}
    \ifx\arg\empty
    {\hfill\vbox{\hsize#2\hrule\hbox to #2{\vrule\hfill\vbox to #1{\hsize#2\vfill}\vrule}\hrule}\hfill}
    \else
    {\hfill\epsfbox{#3}\hfill}
    \fi}

\long\def\inst#1{\par\nobreak\kern 4pt\nobreak
    {\it #1}\par\vskip 10pt plus 3pt minus 3pt}

\begin{document}
\pagenumbering{arabic}
\preprint{\babar-PUB-13/019} 
\preprint{SLAC-PUB-SLAC-PUB-16048} 
\input{authors_oct2013_bad2479.tex}

\begin{flushleft}
\babar-PUB-13/019\\
SLAC-PUB-16048\\

\end{flushleft}
\title{ {\large {\bf \boldmath Study of $B^{\pm,0} \rightarrow J/\psi K^+ K^- K^{\pm,0}$ and search for $B^0 \rightarrow J/\psi \phi$ at \babar}}}

\begin{abstract}
We study the rare $B$ meson decays  $B^{\pm,0} \rightarrow J/\psi K^+ K^- K^{\pm,0}$, $B^{\pm,0} \rightarrow J/\psi \phi K^{\pm,0}$, and search for 
$B^0 \rightarrow J/\psi \phi$,  using 469 million $B \overline B$ events collected at the $\Upsilon(4S)$ resonance with the \babar 
detector at the PEP-II ~$e^+ e^-$ asymmetric-energy collider. We present new measurements of branching fractions and a study of the $J/\psi \phi$
mass distribution in search of new charmonium-like states. In addition, we search for the decay $B^0 \rightarrow J/\psi \phi$, and find no evidence of a signal.
\end{abstract}

\maketitle

\section{Introduction}

Many charmonium-like resonances have been discovered in the past, revealing a spectrum too rich to interpret in terms of conventional mesons expected from potential models~\cite{NRPM}. In several cases, it has not been possible to assign a spin-parity value to the resonance. Some of them have been extensively investigated as possible candidates for non-conventional mesons, such as tetraquarks, glueballs, or hybrids~\cite{XYZrev}. 

In a search for exotic states, the CDF experiment studied the decay $B^+ \rightarrow J/\psi \phi K^+$~\cite{conj}, where $J/\psi \rightarrow \mu^+ \mu^-$ and $\phi(1020) \rightarrow K^+ K^-$, claiming the observation of a resonance labeled the $X(4140)$ decaying to $J/\psi \phi$~\cite{kai}. They found evidence in the same decay mode for another resonance, labeled as the $X(4270)$~\cite{kai-bis}. Recently, the LHCb experiment studied the decay $B^+ \rightarrow J/\psi \phi K^+$ in $pp$ collisions at 7~TeV, with a data sample more than three times larger than that of CDF, and set an upper limit (UL) incompatible with the CDF result~\cite{LHCb}. The D0 and the CMS experiments more recently made studies of the same decay channel, leading to different conclusions~\cite{cms, d0} than the LHCb experiment. In this work we study the rare decays  $B^{+} \rightarrow J/\psi K^+K^- K^{+}$, $B^{0} \rightarrow J/\psi K^+K^- K_S^0$ and search for possible resonant states in the $J/\psi \phi$ mass spectrum. We also search for the decay $B^0 \rightarrow J/\psi \phi$, which is expected to proceed mainly via a Cabibbo-suppressed and color-suppressed transition $\bar b d \rightarrow \bar c c \bar d d$. The absence of a signal would indicate that the required rescattering of $\bar d d$ into $\bar s s$ is very small.

This paper is organized as follows. In Sec. II we describe the detector and data selection and in Sec. III we report the branching-fraction (BF) measurements. Section IV is devoted to the resonance search, while Sec. V summarizes the results.
\section{The \babar detector and data selection}
We make use of the data set collected by the \babar\ detector at the PEP-II $e^{+}e^{-}$ storage rings operating at the $\FourS$ resonance. The integrated luminosity for this analysis is 422.5 fb$^{-1}$, which corresponds to the production of 469 million $B\overline{B}$ pairs~\cite{lumi}.

The \babar\ detector is described in detail elsewhere~\cite{luminew}. We mention here only the components of the detector that are used in the present analysis. Charged particles are detected and their momenta measured with a combination of a cylindrical drift chamber (DCH) and a silicon vertex tracker (SVT), both operating within the 1.5 T magnetic field of a superconducting solenoid. Information from a ring-imaging Cherenkov detector (DIRC) is combined with specific ionization measurements from the SVT and DCH to identify charged kaon and pion candidates. The efficiency for kaon identification is 90\% while the rate for a pion being misidentified as a kaon is 2\%.  For low transverse momentum kaon candidates that do not reach the DIRC, particle identification relies only on the energy loss measurement, so that the transverse momentum  spectrum of identified kaons extends down to 150 \mevc. Electrons are identified using information provided by a CsI(Tl) electromagnetic calorimeter (EMC), in combination with that from the SVT and DCH, while muons are identified in the Instrumented Flux Return (IFR). This is the outermost subdetector, in which muon/pion discrimination is performed. Photons are detected, and their energies measured with the EMC.

For each signal event candidate, we first reconstruct the \jpsi by geometrically constraining to a common vertex a pair of oppositely charged tracks, identified as either electrons or muons, and apply a loose requirement that the $\chi^2$ fit probability exceed 0.1\%. For $\jpsi \rightarrow \ee$ we use bremsstrahlung energy-loss recovery: if an electron-associated photon cluster is found in the EMC, its three-momentum vector is incorporated into the calculation of the invariant mass $m_{e^{+}e^{-}}$. The vertex fit for a \jpsi candidate includes a constraint to the nominal \jpsi mass value~\cite{PDG}.

For $B^+ \rightarrow J/\psi K^+ K^- K^+$ candidates, we combine the $J/\psi$ candidate with three loosely identified kaons and require a vertex-fit probability larger than 0.1\%. Similarly, for $B^0 \rightarrow J/\psi K^- K^+ \KS$ candidates, we combine the $\jpsi$ and $\KS$ with two loosely identified kaons and require a vertex-fit probability larger than 0.1\%.

A $\KS$ candidate is formed by geometrically constraining a pair of oppositely charged tracks to a common vertex, with $\chi^2$ fit probability larger than 0.1\%. The pion mass is assigned to the tracks without particle-identification (PID) requirements. The three-momenta of the two pions are then added and the \KS energy is computed using the nominal \KS  mass. We require the \KS flight length significance with respect to the $B^0$ vertex to be larger than 3$\sigma$.

We further select $B$ meson candidates using the energy difference \DeltaE $\equiv E^{\ast}_B-\sqrt{s}/2$ in the center-of-mass frame and the beam-energy-substituted mass defined as $\mes \equiv \sqrt{((s/2+\vec{p}_i\cdot\vec{p}_B)/E_i)^2-\vec{p}_B^{\,2}}$, where ($E_i,\vec{p}_i$) is the initial state \ee four-momentum vector in the laboratory frame and $\sqrt{s}$ is the center-of-mass energy. In the above expressions $E^{\ast}_B$ is the $B$ meson candidate energy in the center-of-mass frame, and $\vec{p}_B$ is its laboratory frame momentum.

When multiple candidates are present, the combination with the smallest \DeltaE is chosen. We find that, after requiring \mes$>5.2$ \gevcc, the fraction of events having multiple candidates is 1.3\% for $B^+$ and 8.6\% for $B^0$. From simulation, we find that 99.6$\%$ of the time we choose the correct candidate.

The final selection requires $\left | \DeltaE \right|<30$ \mev and $\left | \DeltaE\right|<25$ \mev for $B^+$ and $B^0$ decays, respectively; the additional selection criterion \mes$>5.2$ \gevcc is required for the calculation of the BFs, while \mes$>5.27$ \gevcc is applied to select the signal region for the analysis of the invariant mass systems.

\section{Branching Fractions}
\begin{figure*}[ht] 
\begin{center}
\mbox{
\scalebox{0.28}{\includegraphics{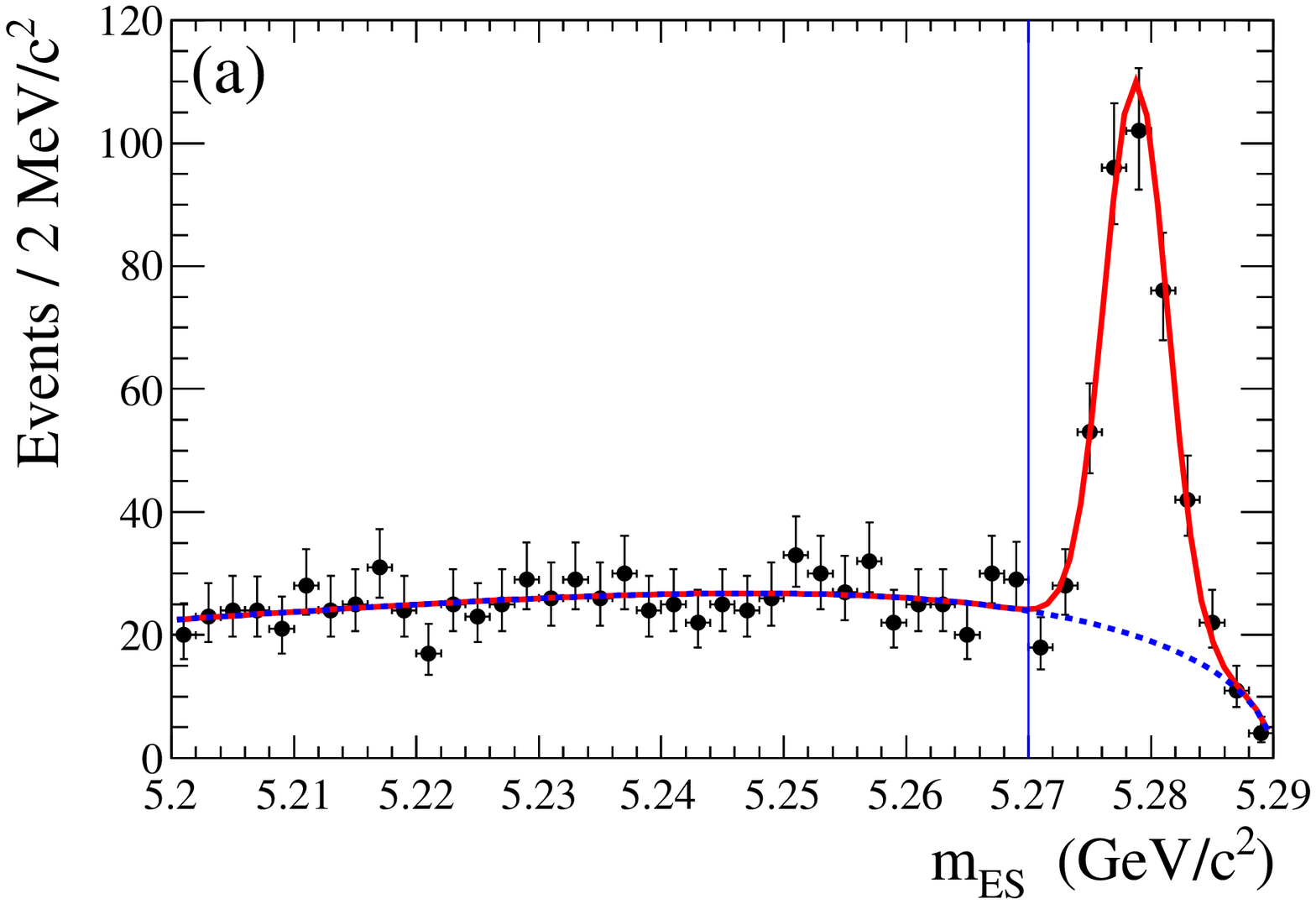}}
\scalebox{0.28}{\includegraphics{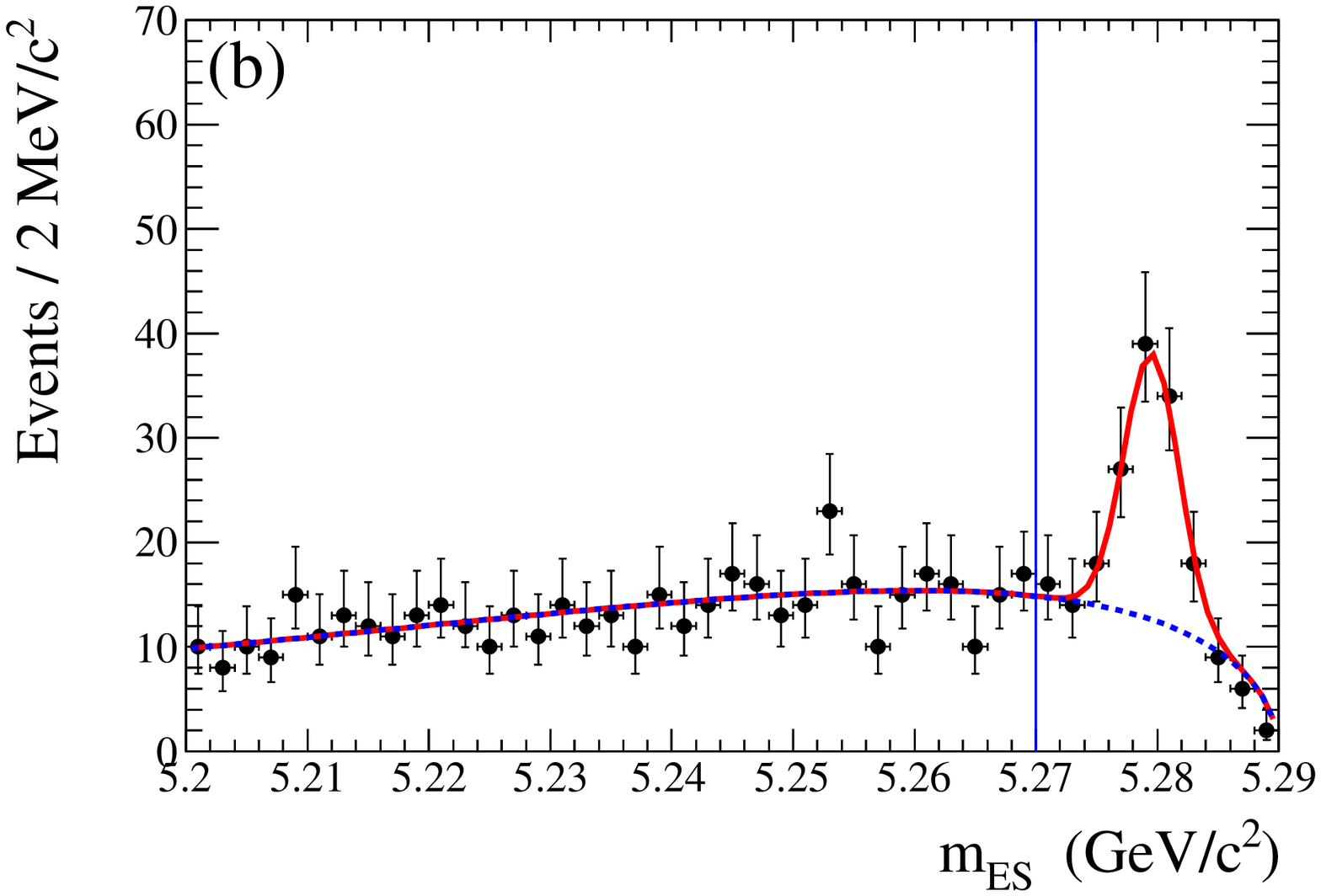}}}
\mbox{
\scalebox{0.28}{\includegraphics{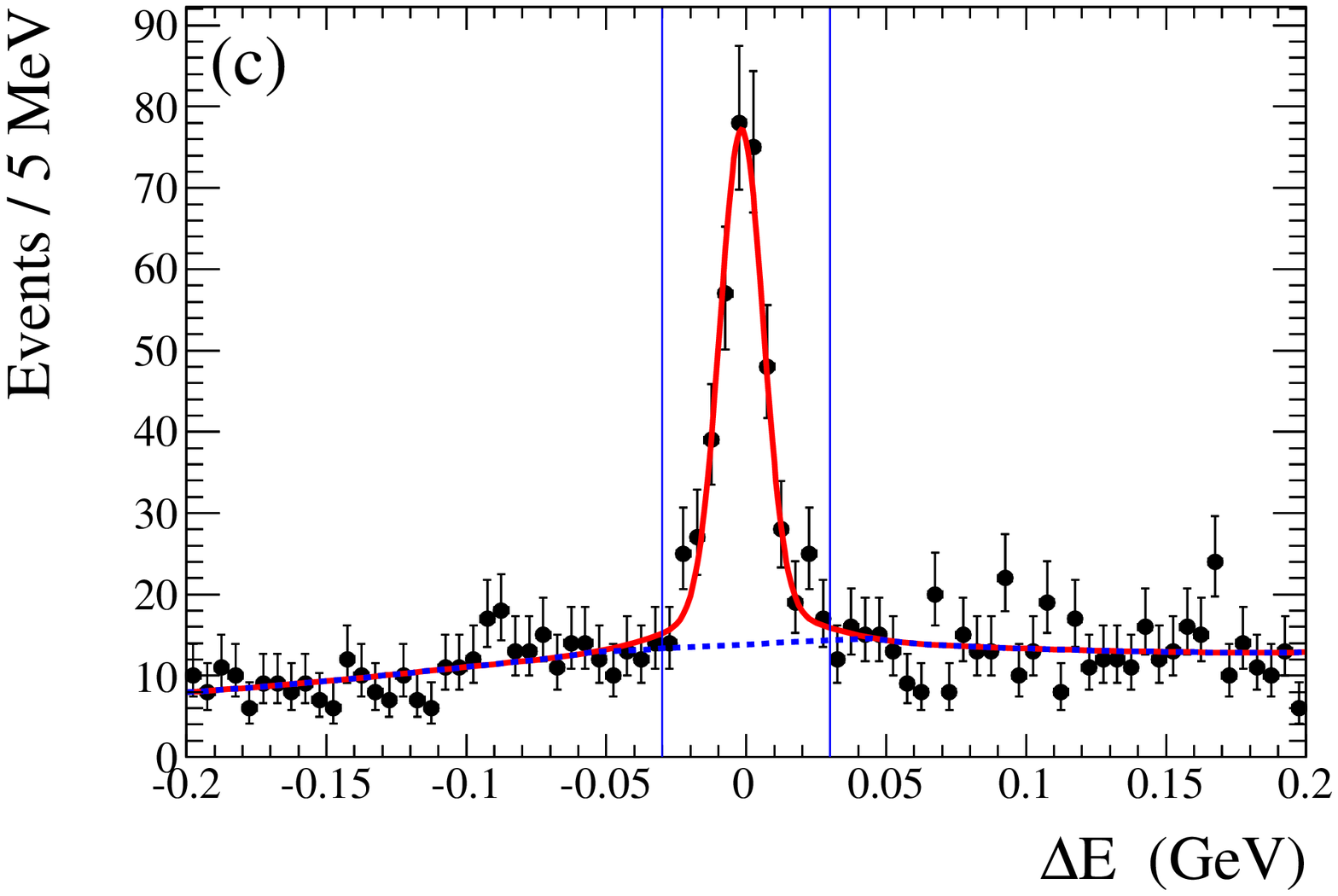}} 
\scalebox{0.28}{\includegraphics{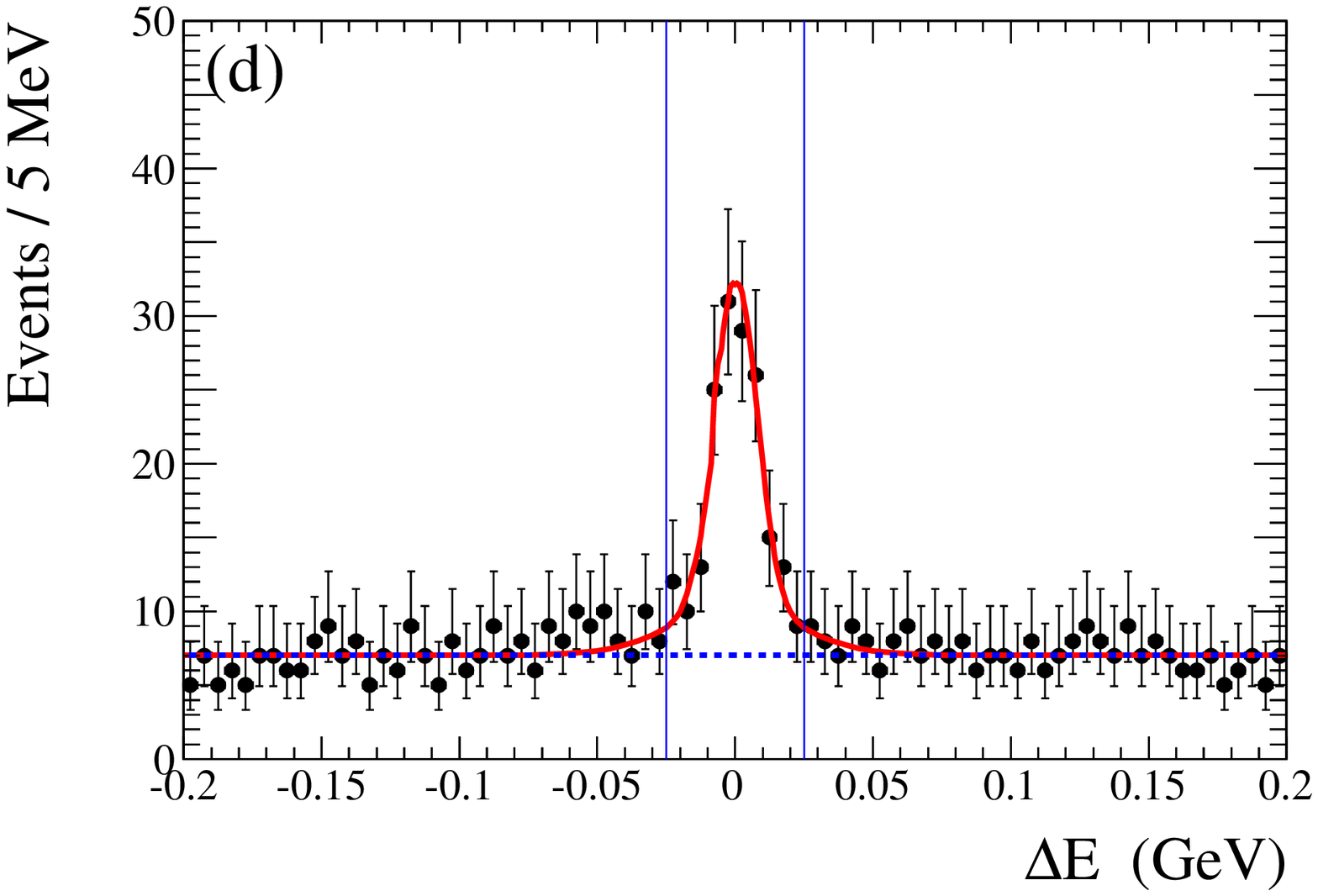}}}
\caption{\label{fig1} The \mes distributions for (a) $B^+ \rightarrow J/\psi K^+ K^- K^+$ and (b) $B^0 \rightarrow J/\psi  K^- K^+ \KS$, for the \DeltaE regions indicated in the text. The \DeltaE distributions for \mes$>5.27$ \gevcc are shown for (c) $B^+ \rightarrow J/\psi K^+ K^- K^+$ and (d) $B^0 \rightarrow J/\psi  K^- K^+ \KS$. The continuous (red) curve represents the signal plus background, while the dotted (blue) curve represents the fitted background. Vertical (blue) lines indicate the selected signal regions.}
\end{center}
\end{figure*}
\begin{figure*}[ht] 
\begin{center}
\mbox{
\scalebox{0.28}{\includegraphics{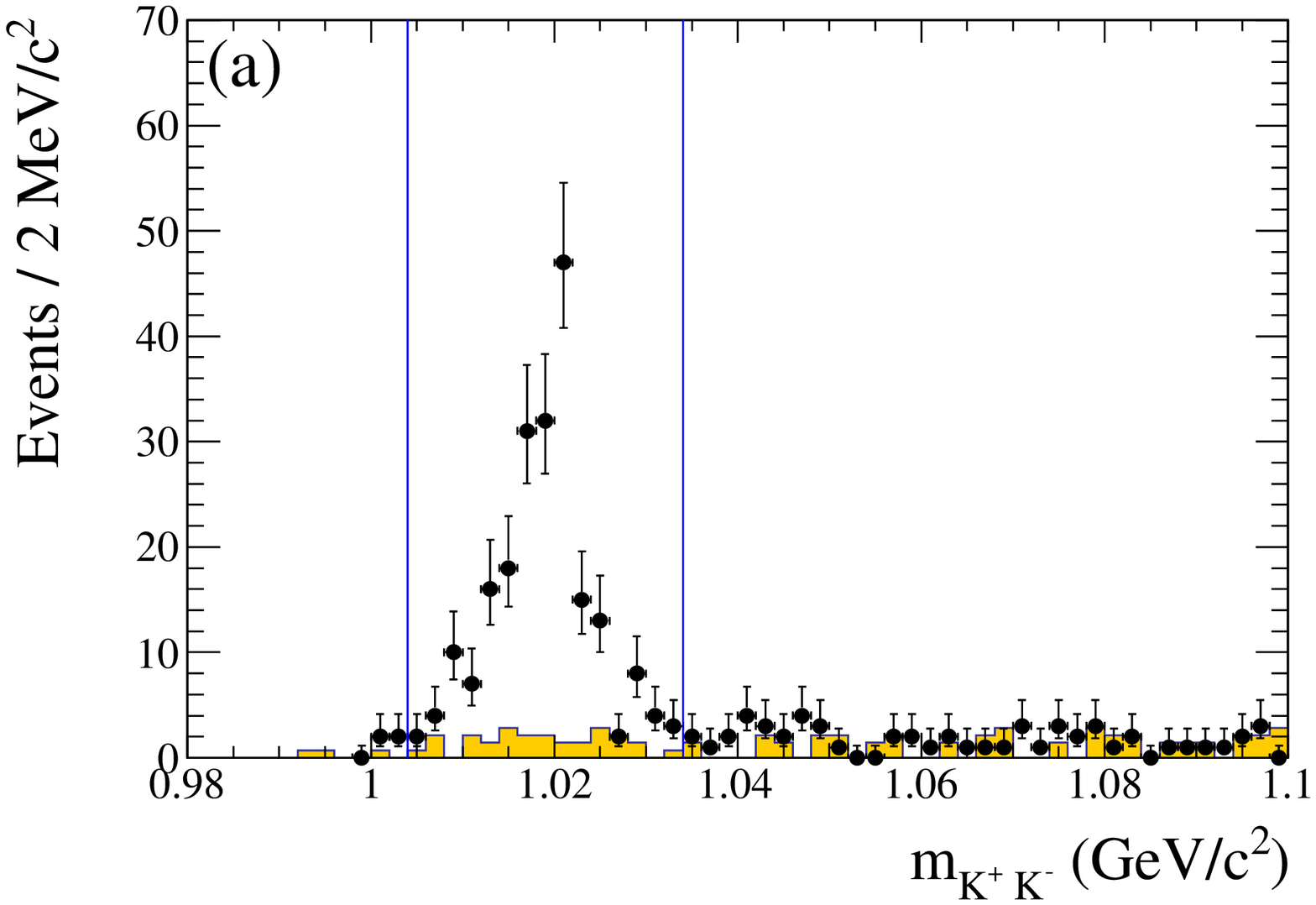}}
\scalebox{0.28}{\includegraphics{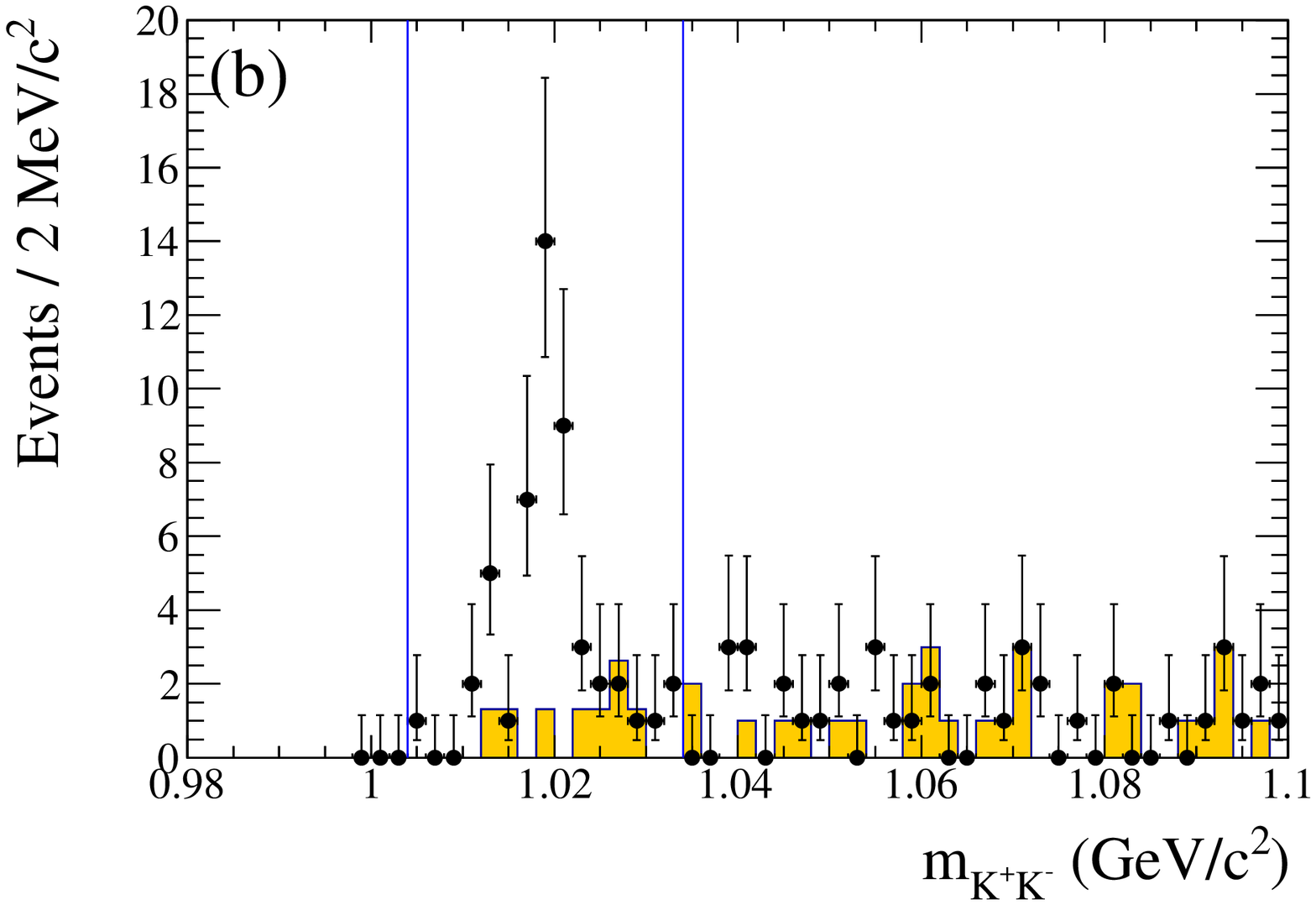}}
}
\mbox{
\scalebox{0.28}{\includegraphics{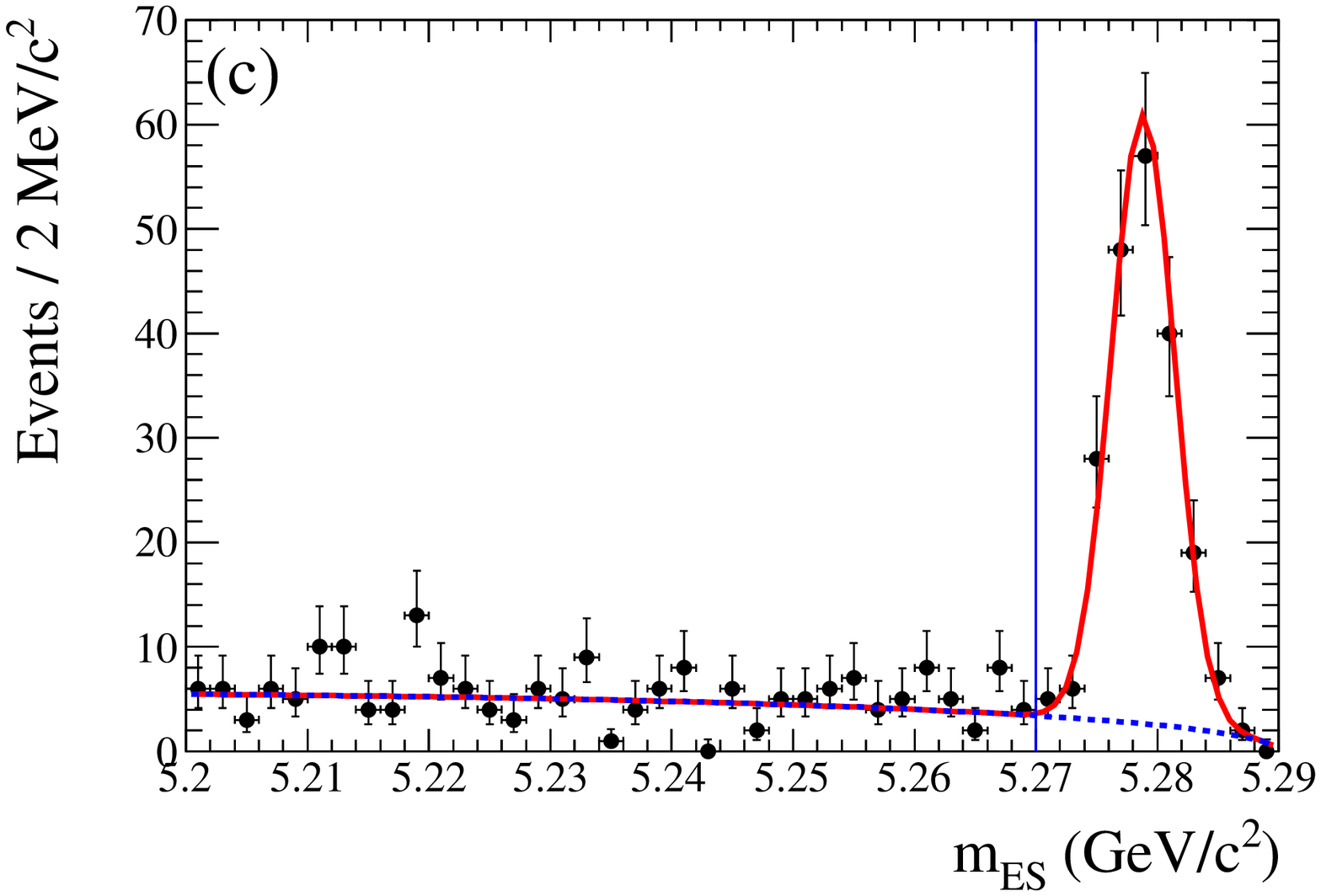}}
\scalebox{0.28}{\includegraphics{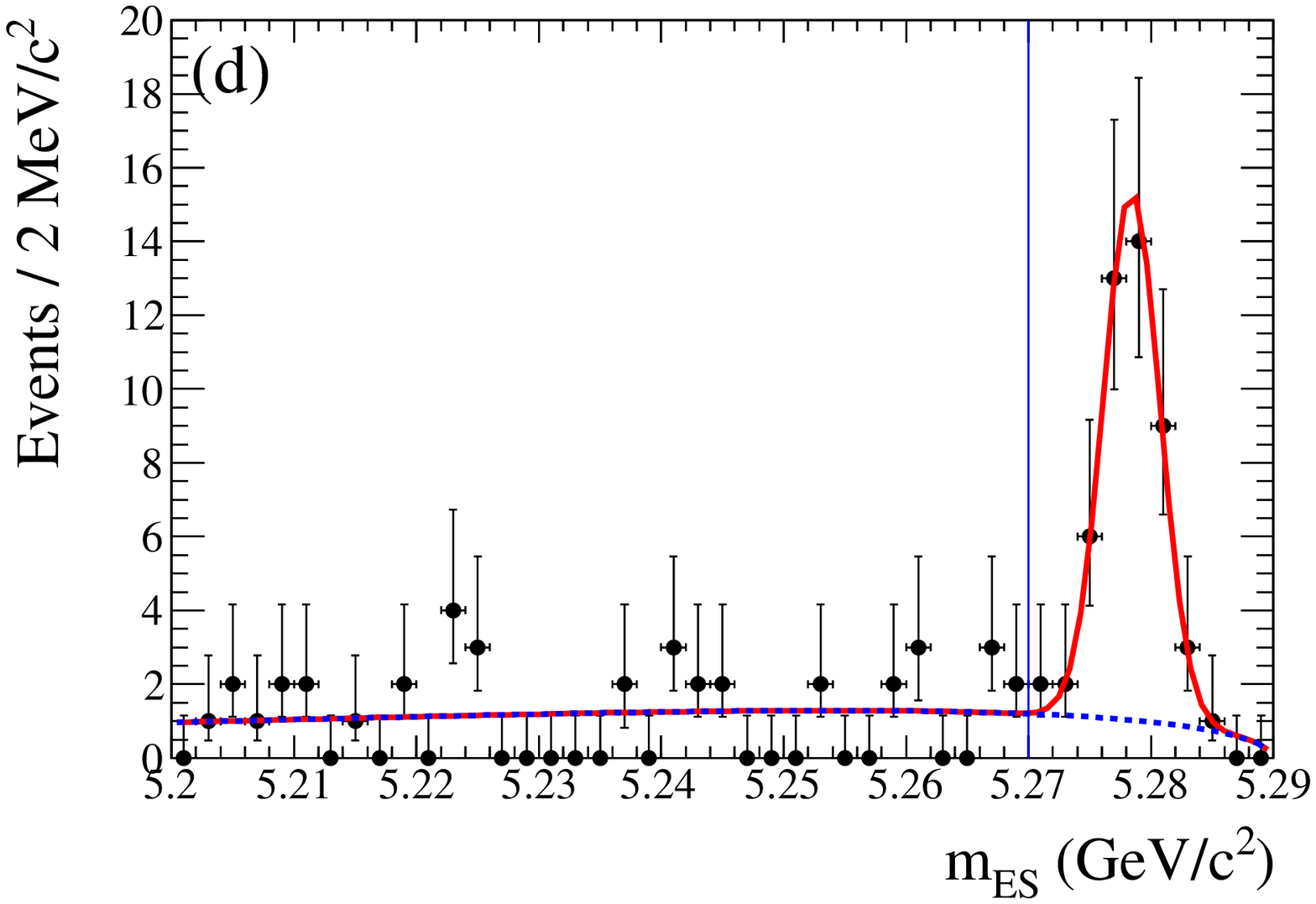}}
}
\mbox{
\scalebox{0.28}{\includegraphics{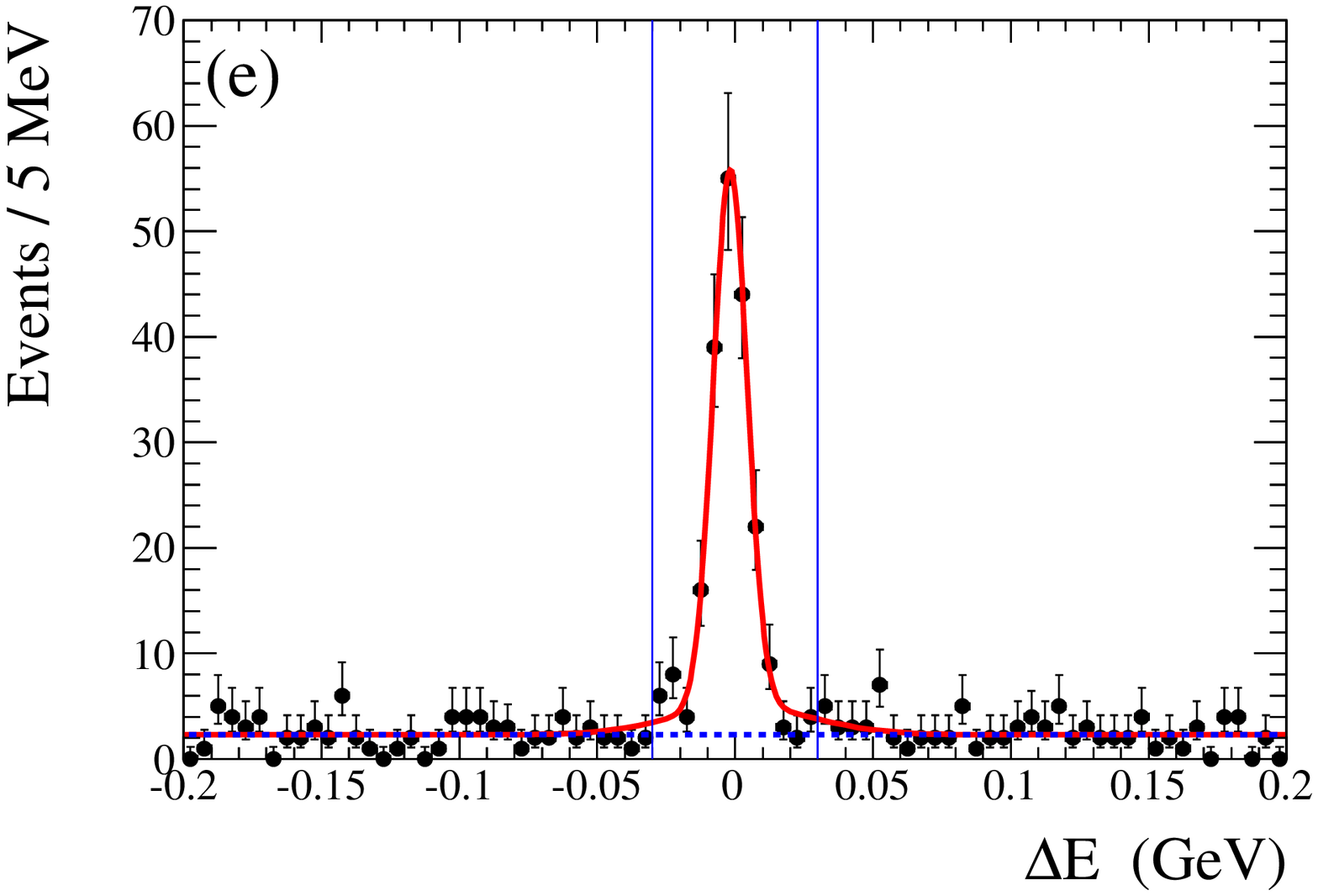}}
\scalebox{0.28}{\includegraphics{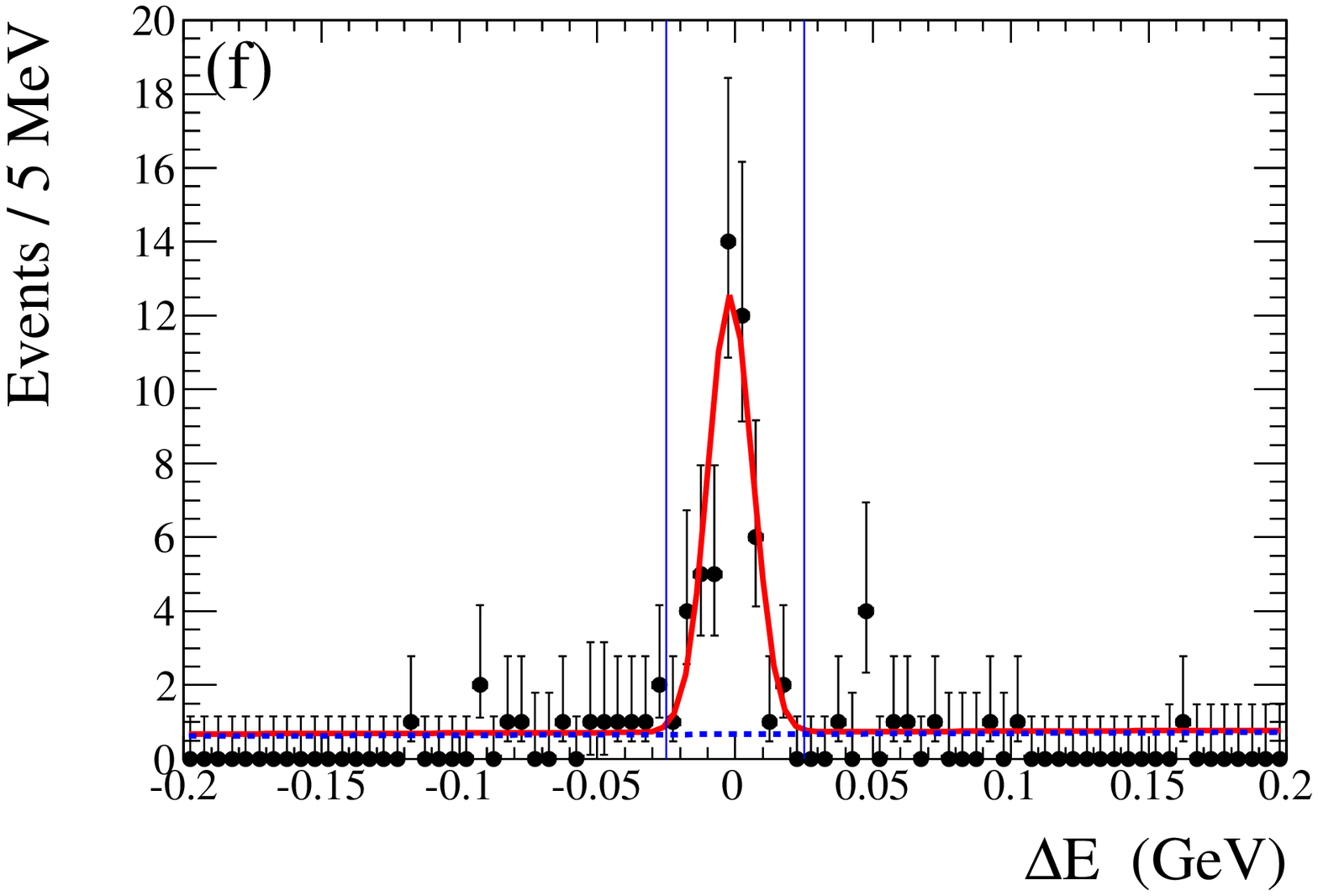}}}

\caption{\label{fig2} (a) The $K^+K^-$ mass spectrum,  (c) \mes, and (e) \DeltaE distribution for $B^+ \rightarrow J/\psi \phi K^+$. (b) The $K^+K^-$ mass spectrum,  (d) \mes, and (f) \DeltaE distribution for $B^{0} \rightarrow J/\psi  \phi \KS$. The dots are the data points, the shaded (yellow) distributions are obtained from the \DeltaE sidebands. Vertical (blue) lines indicate the selected signal regions. In (a) and (b) the \mes and \DeltaE selection criteria described in Sec. II have been applied.}
\end{center}
\end{figure*}
\begin{figure*}[ht] 
\begin{center}
\mbox{
\scalebox{0.28}{\includegraphics{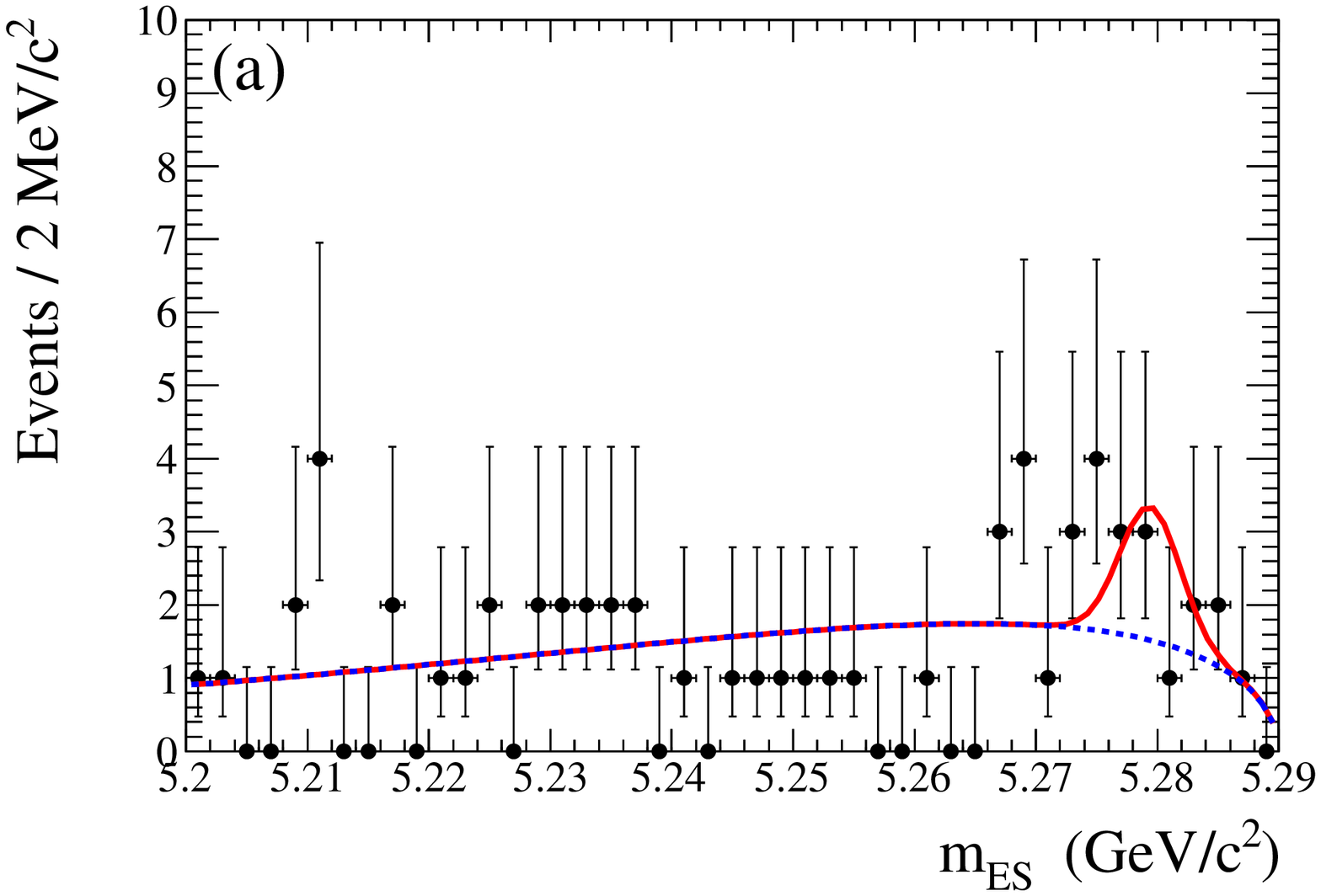}}
\scalebox{0.28}{\includegraphics{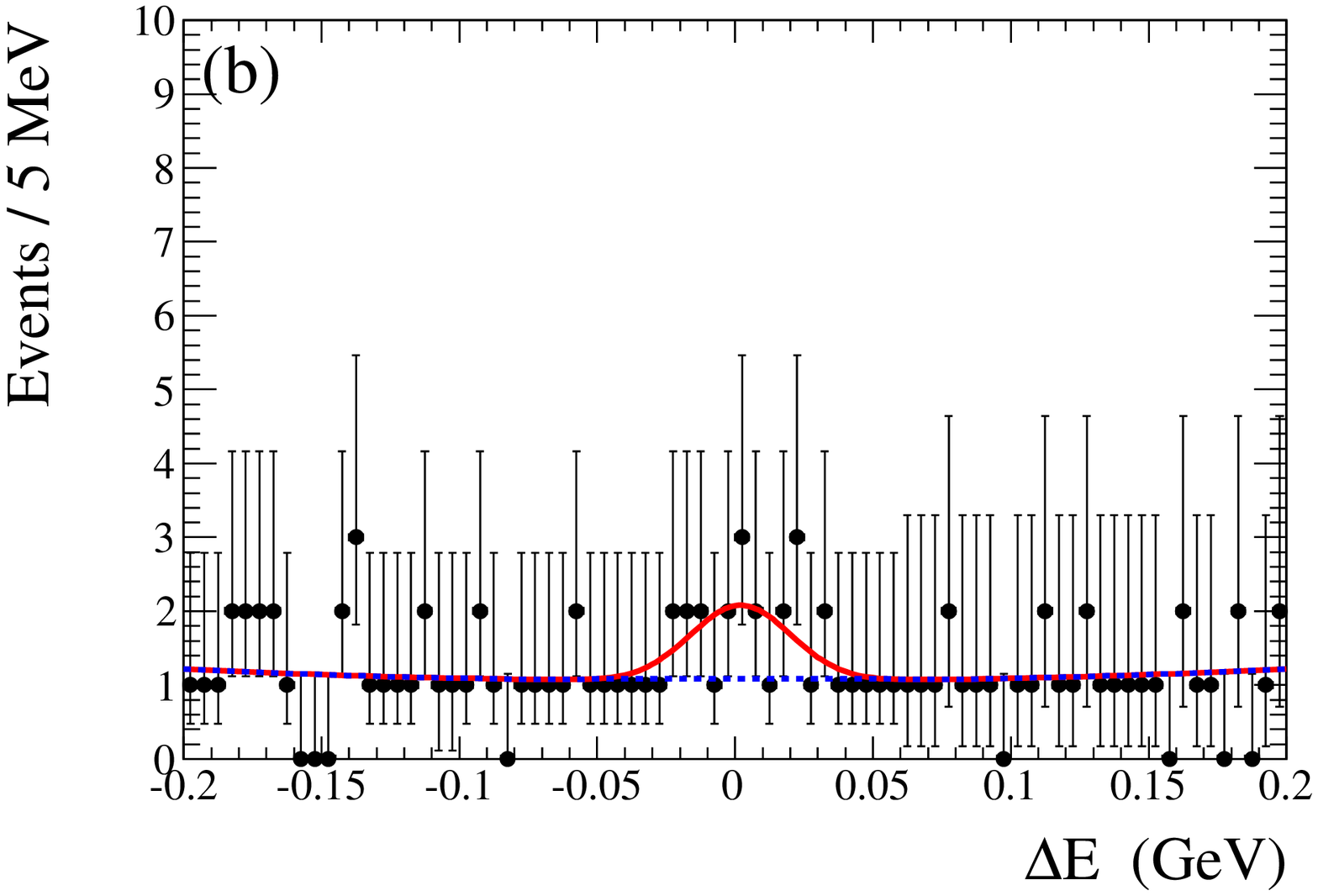}}
}
\caption{\label{fig3} (a) The \mes  and (b) \DeltaE  distribution for $B^0 \rightarrow J/\psi \phi$ event candidates. The curves in (a) and (b) are the result of the fits described in the text.}
\end{center}
\end{figure*}
\begin{table*}[!htb]
\caption{Event yields, efficiencies ($\epsilon$) and  BF measurements (\BR) for the different decay modes. For channels involving \KS,  the yields and efficiencies refer to $\KS \to \pip \pim$, the BF includes the corrections for $K^0_S \rightarrow \pi^0 \pi^0$ and $K^0_L$ decay. The $B^0 \rightarrow J/\psi \phi$ UL at 90$\%$ c.l. is listed at the end of the table.}
\begin{center}
\begin{tabular}{lrcccccl}
\hline\hline
$B$ channel & Event& $\epsilon$ ($\%$)& Corrected                            &\BR\ ($\times 10^{-5}$) \\
          & yield&                       & yield    &  \\
\hline 
\noalign{\vskip2pt}
$B^+ \rightarrow J/\psi K^+ K^- K^+$    &$~~$290$\pm$22&$~~~$15.08$\pm$0.04&~~1923$\pm$146~&~~3.37$\pm$0.25$\pm$0.14 \cr
$B^+ \rightarrow J/\psi \phi K^+$   &$~~$189$\pm$14&$~~~$13.54$\pm$0.04 &~~1396$\pm$103  &~~5.00$\pm$0.37$\pm$0.15\cr
$B^0 \rightarrow J/\psi K^+ K^- K^0$   &$~~$68$\pm$13 &$~~~$10.35$\pm$0.04 &~~~~657$\pm$126&~~3.49$\pm$0.67$\pm$0.15\cr
$B^0 \rightarrow J/\psi \phi K^0$ &41$\pm$~7     &$~~~$10.10$\pm$0.04 &~~~406$\pm$69& ~~4.43$\pm$0.76$\pm$0.19\cr
$B^0  \rightarrow J/\psi \phi$   &~~~~6 $\pm$ 4  &$~~~$31.12$\pm$0.07 &~~~~19~$\pm$13&~$<0.101$      \cr
\hline
\end{tabular}
\end{center}
\label{tab:tab1}
\end{table*}
\begin{table*}[!htb]
\caption{Systematic uncertainty contributions ($\%$) to the evaluation of the BFs.}
\begin{center}
\begin{tabular}{lcccccccc}
\hline\hline
 Source&$B^+ \rightarrow J/\psi K^+ K^- K^+$ ~&  ~~$B^+ \rightarrow J/\psi \phi K^+$~ &~~$B^0 \rightarrow J/\psi  K^- K^+  \KS$~&~~$B^0 \rightarrow J/\psi \phi \KS$~&~~$B^0 \rightarrow J/\psi \phi$\\
\hline  
\noalign{\vskip2pt}
$B \overline B$  counting  &  0.6  &   0.6   &  0.6  & 0.6  & 0.6\cr
Efficiency    &  0.04 &   0.04  &  0.04 & 0.04 & 0.07\cr
Tracking      & 0.9  & 0.9   & 1.2    &1.2 & 0.7\cr
\KS       &   $-$   &   $-$  &    1.7    & 1.7 &  $-$\cr
Secondary BFs  & 0.08   &0.5      &0.1      &0.5  & 0.5\cr
Decay model         &  $-$   &0.4       & $-$        &0.9 & 1.0\cr
$\rm pdfs$       & 3.0  &0.7      &2.0      & 0.5   &  1.0 \cr
PID           & 2.5   &2.5      &3.0     &3.0  & 2.0 \cr 
\hline
Total contribution&4.1&3.0&4.2&4.4&2.7 \cr
\hline
\end{tabular}
\end{center}
\label{tab2}
\end{table*}
\begin{figure*}[ht] 
\begin{center}
\mbox{
\scalebox{0.28}{\includegraphics{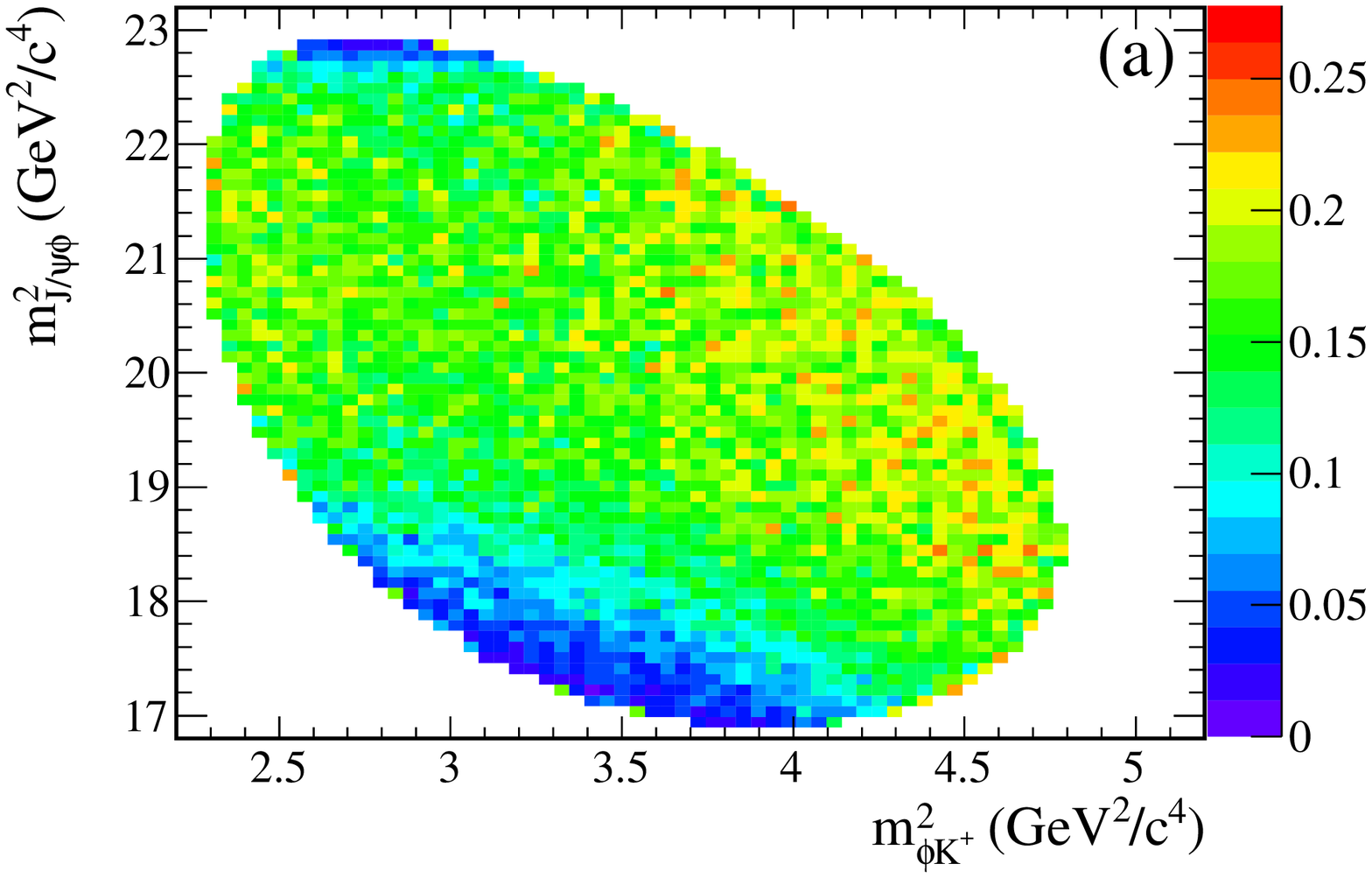}}  \hspace{5mm}
\scalebox{0.30}{\includegraphics{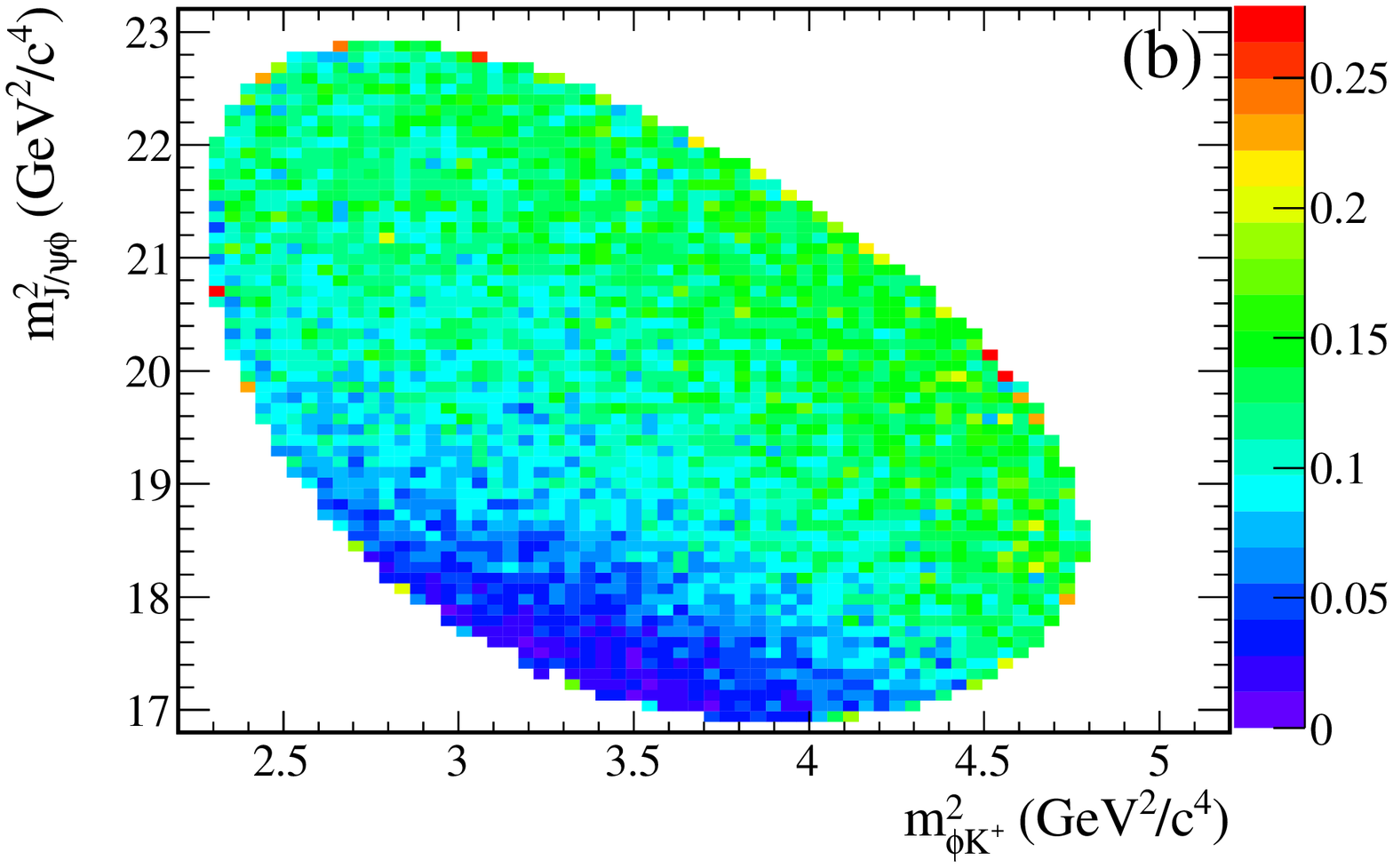}}  \hspace{-5mm}
}
\caption{\label{fig5} Efficiency distribution on the Dalitz plot for (a) $B^+ \rightarrow J/\psi \phi K^+$ and (b) $B^0 \rightarrow J/\psi \phi \KS$.} 
\end{center}
\end{figure*}
\begin{figure*}[ht] 
\begin{center}
\mbox{
\scalebox{0.28}{\includegraphics{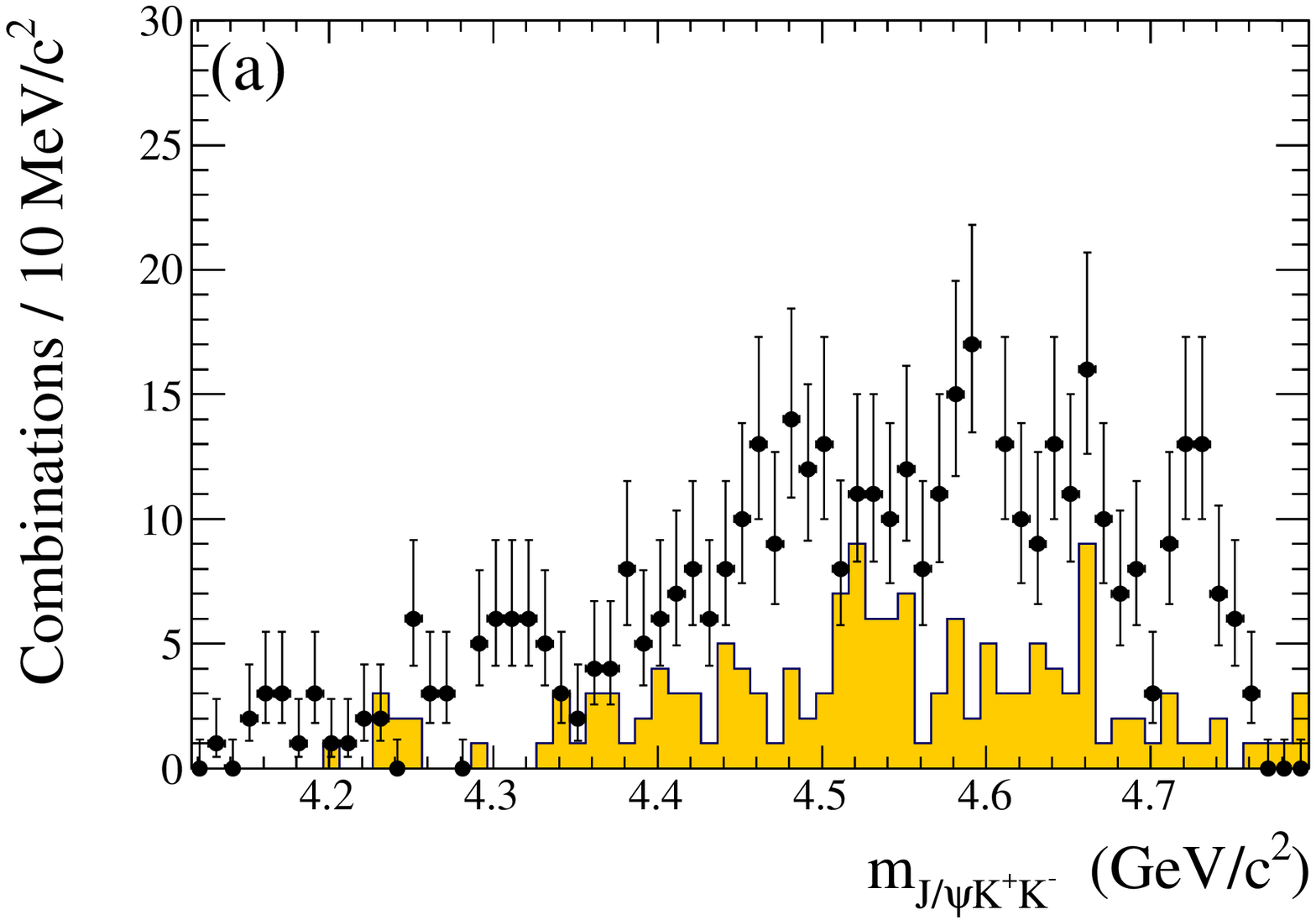}}
\scalebox{0.28}{\includegraphics{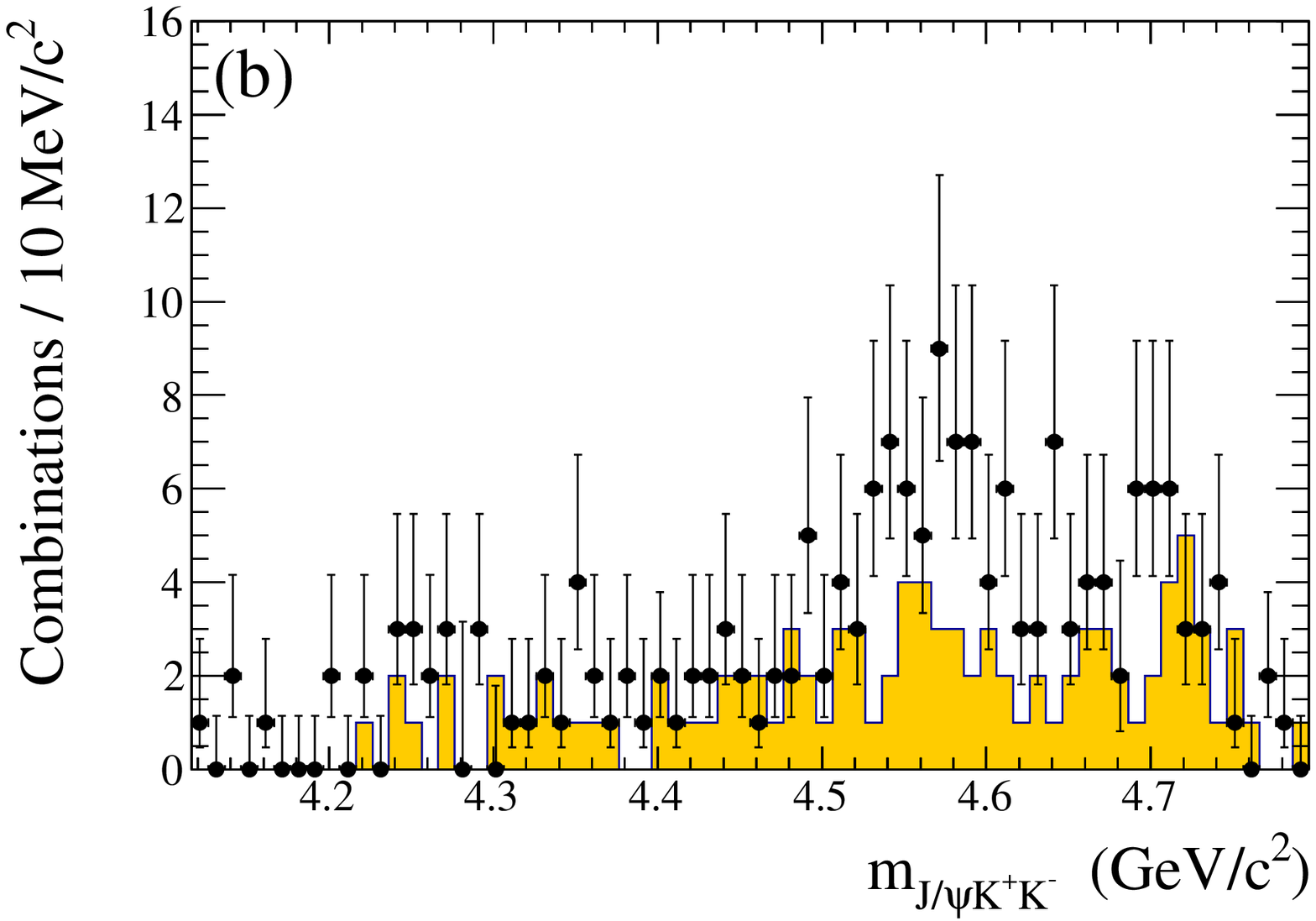}}
}
\caption{\label{fig4} Invariant mass distribution $J/\psi K^+ K^- $ for (a) $B^+ \rightarrow J/\psi K^+ K^- K^+$ and (b) $B^0 \rightarrow J/\psi K^+ K^- \KS$.
The shaded (yellow) histogram on each figure indicates the background estimated from the \DeltaE~ sidebands.}  
\end{center}
\end{figure*}

Figure~\ref{fig1} ~shows ~the ~\mes ~~distributions ~for ~(a) $B^+ \rightarrow J/\psi K^+ K^- K^+$ and (b) $B^0 \rightarrow J/\psi K^- K^+ \KS$ candidates after having applied the \DeltaE selections described in Sec.~II, while the corresponding \DeltaE distributions are shown in Fig.~\ref{fig1}(c) and  Fig.~\ref{fig1}(d), respectively, for \mes$>5.27$ \gevcc. Figure~\ref{fig2} shows the $K^+K^-$ invariant mass distribution in the region $m_{K^+ K^-} <$1.1 \gevcc for (a) $B^+$ and (b) $B^0$ candidates. A clean $\phi(1020)$ signal is present in both mass spectra. The background contributions, estimated from the $\DeltaE$ sidebands in the range $\rm{40}<|\DeltaE|<\rm{70~ MeV}$, are shown as shaded histograms in Fig.~\ref{fig2}(a) and Fig.~\ref{fig2}(b) and are seen to be small. In the following we have ignored the presence of possible additional S-wave contributions in the $\phi(1020)$ signal region.

We select the $\phi(1020)$ signal region to be in the mass range [1.004$-$1.034] \gevcc.
Figure~\ref{fig2} shows the \mes distribution for (c) $B^+ \rightarrow J/\psi \phi K^+$ and (d) $B^0 \rightarrow J/\psi \phi \KS$ candidates, respectively, for events in the $\phi$ mass region, which satisfy the \DeltaE selection criteria. 
Figures~\ref{fig2}(e) and~\ref{fig2}(f) show the \DeltaE distribution for \mes$>5.27$ \gevcc, when requiring the $\Kp \Km$ invariant mass to be in the $\phi(1020)$ signal region. The distributions of Fig.~\ref{fig2}(c) and Fig.~\ref{fig2}(e) contain 212 events in the \mes and $\DeltaE$ signal region, with an estimated background of 23 events. Similarly, those of  Fig.~\ref{fig2}(d) and Fig.~\ref{fig2}(f) contain 50 events, with an estimated background of 9 events.

We search for the decay $B^0 \rightarrow J/\psi \phi$ by constraining  a fitted \jpsi and two loosely identified kaon candidates to a common vertex.  Possible backgrounds originating from the decay $B^0 \rightarrow J/\psi K^{0*}(892)$, $K^{0*}(892) \rightarrow K^- \pi^+$, and from the channel $B^0 \rightarrow J/\psi K_1(1270)$, $K_1(1270) \rightarrow K^- \pi^+ \pi^{0}$ are found consistent with zero, after applying a dedicated selection as described in Sec.~II and Sec.~III.  Figure~\ref{fig3} shows the corresponding \mes and \DeltaE distributions. We do not observe a significant signal for this decay mode. 

For Figs. 1-3 an unbinned maximum likelihood fit to each \mes distribution is performed to determine the yield and obtain a BF measurement~\cite{formula}.
We use the sum of two functions to parametrize the \mes distribution; a Gaussian function describes the signal, and an ARGUS function~\cite{argus} the background. A study of the \DeltaE~sidebands  did not show the presence of peaking backgrounds. Table~\ref{tab:tab1} summarizes the fitted yields obtained.
 
As a validation test, we fit the \DeltaE distributions shown in Figs. 1-3, using a double-Gaussian model for the signal and a linear function for the background, and we obtain yields consistent with those from the \mes fits. 

The signals in Fig.~\ref{fig1}, corresponding to the $B^+ \rightarrow J/\psi K^+ K^- K^+$ and the $B^0 \rightarrow J/\psi K^+ K^- \KS$  decays, yield 14.4$\sigma$ and 5.5$\sigma$ significance, respectively. Those in Fig.~\ref{fig2}, which restrict the invariant mass $m_{K^+K^-}$ to the signal region of the $\phi(1020)$ meson, are observed with significance 16.1$\sigma$ and 5.6$\sigma$, respectively. In this paper the statistical significance of the peaks is evaluated as $\sqrt{-2  ln(L_{0}/L_{\rm max})}$, where $L_{\rm max}$ and $L_{0}$ represent the maximum likelihood values with the fitted signal yield and with the signal yield fixed to zero, respectively.

We estimate the efficiency for the different channels using Monte Carlo (MC) simulations. For each channel we perform full detector simulations where $B$ mesons decay uniformly over the available phase space (PHSP). These simulated events are then reconstructed and analyzed as are the real data. These MC simulations are also used to validate the analysis procedure and the BF extractions.  

Table~\ref{tab:tab1} reports the resulting integrated efficiencies for the 
different channels, and the efficiency-corrected yields. The efficiency is computed in two different ways. For $B^{+} \rightarrow J/\psi \phi K^{+}$ and $B^{0} \rightarrow J/\psi \phi \KS$ we make use of a 
Dalitz-plot-dependent efficiency, where each event is weighted by the inverse of the efficiency evaluated in the appropriate cell of the Dalitz plot shown in Fig.~\ref{fig5}. This approach is particularly important because of the lower efficiency observed at low $J/\psi \phi$ invariant mass, where the spectrum deviates from pure PHSP behavior. For the $\phi$ channels, the ``Corrected yield'' values in Table I are obtained as sums of inverse Dalitz-plot efficiencies for events in the $\phi$ signal regions with background-subtraction taken into account as described in Sec. IV.  The events in the $\phi$ signal region account for about 65$\%$ of the data in the four-body final states. There is no evidence of structure in the remaining $\sim$35$\%$ of these events, and so they are corrected according to their average efficiency obtained from MC simulation of four-body PHSP samples. For these channels, $B^{+} \rightarrow J/\psi K^+ K^- K^{+}$ and $B^{0} \rightarrow J/\psi K^+ K^- \KS$, the PHSP corrected yield is added to the $\phi$ signal region corrected yield to obtain the ``Corrected yield'' values in lines 1 and 3 of Table I. The efficiency values in the third column of Table I correspond to ``Event yield'' divided by ``Corrected yield''.

Systematic uncertainties affecting the BF measurements are listed in Table~\ref{tab2}. The evaluation of the integrated luminosity 
is performed using the method of $B \overline B$ counting~\cite{luminew}, and we assign a uniform 0.6\% uncertainty to all the final states. The uncertainty on the efficiency evaluation related to the size of the MC simulations is negligible with respect to the other contributions. The systematic uncertainty on the reconstruction efficiency of charged-particle tracks is estimated from the comparison of data samples and full detector simulations for well-chosen decay modes.  In a similar way we obtain a 1.7\% systematic uncertainty in the reconstruction of \KS meson decays. In the case of the $B^0 \rightarrow \jpsi \phi K_S^{0}$ and  $B^{+} \rightarrow \jpsi \phi K^{+}$ decay modes,  since the \jpsi and the $\phi$ are vector states, we compute the efficiency also under the assumption that the two vector mesons are transversely or longitudinally polarized. We consider the uncertainties related to the choice of the probability density functions ($\rm pdf$) in the fit procedure, by varying fixed parameters by $\pm$1$\sigma$ in their uncertainties. We also evaluate the efficiency variations for different charged-particle-track PID.  All uncertainties are added in quadrature. We note that the BF for $B^+ \rightarrow J/\psi \phi K^+$ and that for $B^0 \rightarrow J/\psi \phi K^0$ are in agreement with their previous \babar measurements~\cite{oldpaper}, which already dominate the PDG average values~\cite{PDG}, but now we obtain more than four times better precision. The combination of these decay modes was observed first by the CLEO Collaboration~\cite{cleo}. Our BF value for the decay $B^+ \rightarrow J/\psi K^+ K^- K^+$ is the first measurement. For the decay $B^0 \rightarrow J/\psi K^+ K^- K^{0}$, the LHCb Collaboration has obtained a BF value ($2.02 \pm 0.43 \pm 0.17 \pm 0.08$)$\times 10^{-5}$~\cite{lhcb_3k}, which is consistent with our result.

We estimate an upper limit (UL) at 90$\%$ confidence level (c.l.) for the BF of the decay $B^0 \rightarrow J/\psi \phi$. The signal yield obtained from the fit to the \mes distribution is 6$\pm$4 events (Fig.~\ref{fig3}(a)), corresponding to an UL at 90$\%$ c.l. of 14 events. The Feldman-Cousins method~\cite{FC} is used to evaluate ULs on BFs. Ensembles of pseudo-experiments are generated according to the $\rm pdfs$ for a given signal yield (10000 sets of signal and background events), and fits are performed. We obtain an UL on the $B^0\to J/\psi\phi$ BF of 1.01$\times$10$^{-6}$. The Belle Collaboration reported a limit of $0.94 \times 10^{-6}$~\cite{belle_jpsiphi}, while a recent analysis from the LHCb Collaboration lowers this limit to $1.9 \times 10^{-7}$~\cite{lhcb_jpsiphi}. 

We compute the ratios
\begin{equation}
R_{+} = \frac{{\cal B}(B^+ \rightarrow J/\psi K^+ K^- K^+)}{{\cal B}(B^+ \rightarrow J/\psi \phi K^+)} = 0.67  \pm  0.07  \pm  0.03
\end{equation}
 and 
\begin{equation}
R_0 = \frac{{\cal B}(B^0 \rightarrow J/\psi K^+ K^- K^0)}{{\cal B}(B^0 \rightarrow J/\psi \phi K^0)} = 0.79 \pm 0.20 \pm 0.05 ,
\end{equation}
and they are consistent with being equal within the uncertainties. We also compute the ratios
\begin{equation} 
R_{\phi} = \frac{{\cal B}(B^0 \rightarrow J/\psi \phi K^0)}{{\cal B}(B^+ \rightarrow J/\psi \phi K^+)} = 0.89 \pm 0.17 \pm 0.04
\end{equation}
and 
\begin{equation} 
R_{2K} = \frac{{\cal B}(B^0 \rightarrow J/\psi K^+ K^- K^0)}{{\cal B}(B^+ \rightarrow J/\psi K^+ K^- K^+)} = 1.04 \pm 0.21 \pm 0.06. 
\end{equation} 

On the basis of the simplest relevant color-suppressed spectator quark model diagrams (e.g. Fig.1 of Ref.~\cite{cleo}), it would be expected that $R_+ = R_0$ and $R_{\phi} \sim R_{2K} \sim$ 1. Our measured values of these ratios are consistent with these expectations.

\begin{table*}
\caption{Results of the fits to the $B \to  \jpsi \phi K$ Dalitz plots. For each fit, the table gives the fit fraction for each resonance, and the 2D and 1D $\chi^2$ values. The fractions are corrected for the background component.}
\label{tab:tab3}
\begin{center}
\vskip -0.2cm
\begin{tabular}{llcccc}
\hline
Channel & Fit & $f_{X(4140)}$(\%) & $f_{X(4270)}$(\%) & 2D $\chi^2/\nu$ & 1D
$\chi^2/\nu$ \cr
\hline
$B^+$ & A & 9.2 $\pm$ 3.3 & 10.6 $\pm$ 4.8 & 12.7/12 & 6.5/20 \cr
 & B & 9.2 $\pm$ 2.9 & 0. & 17.4/13 & 15.0/17 \cr
 & C & 0. & 10.0 $\pm$ 4.8 & 20.7/13 & 19.3/19 \cr
 & D & 0. & 0. & 26.4/14 & 34.2/18 \cr
\hline
$B^0 + B^+ $ & A & 7.3 $\pm$ 3.8 & 12.0 $\pm$ 4.9 & 8.5/12 & 15.9/19 \cr
\hline
\end{tabular}
\end{center}
\end{table*}

\section{Search for Resonance Production}
\begin{figure*}[ht] 
\begin{center}
\mbox{
\scalebox{0.28}{\includegraphics{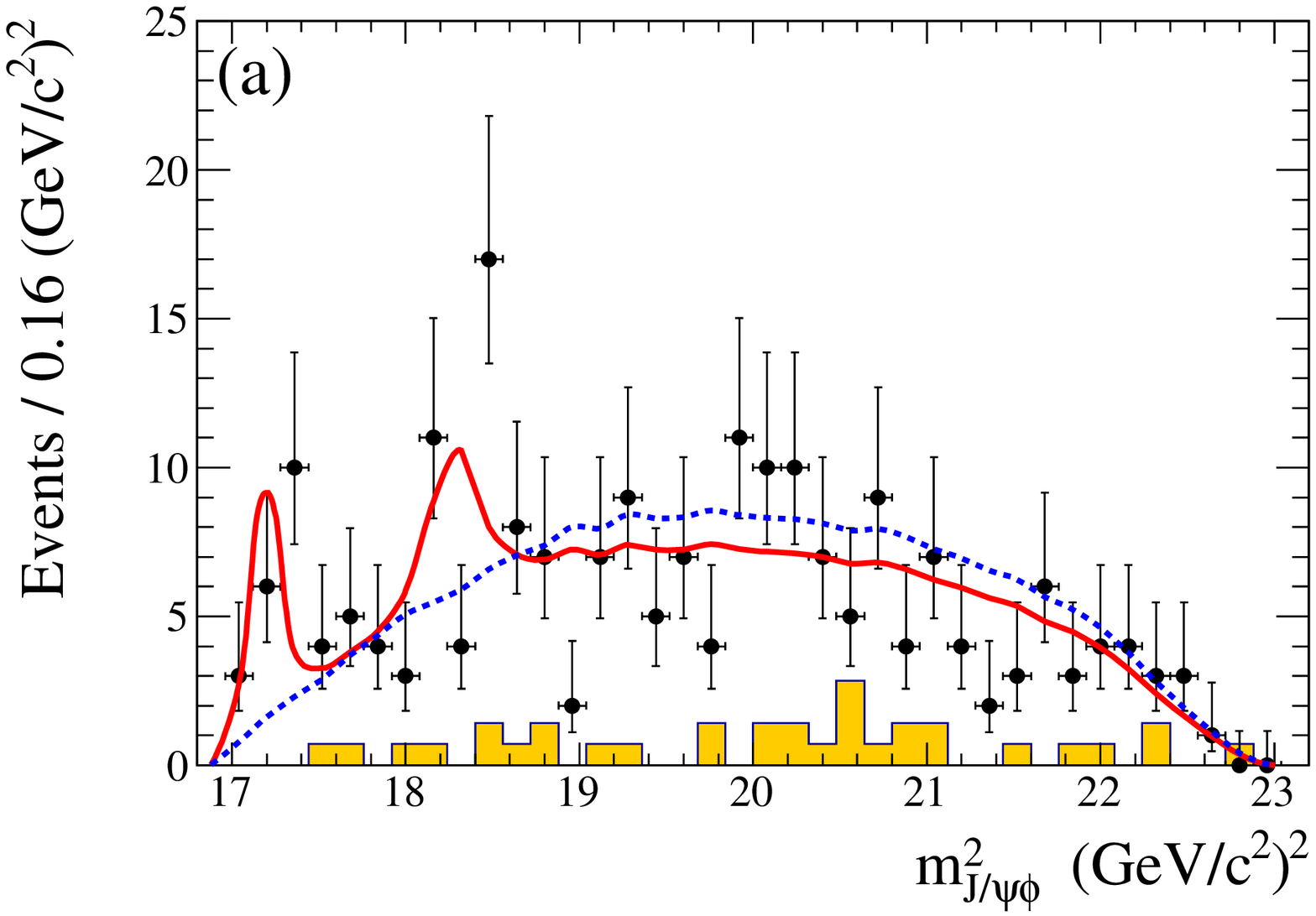}}
\scalebox{0.28}{\includegraphics{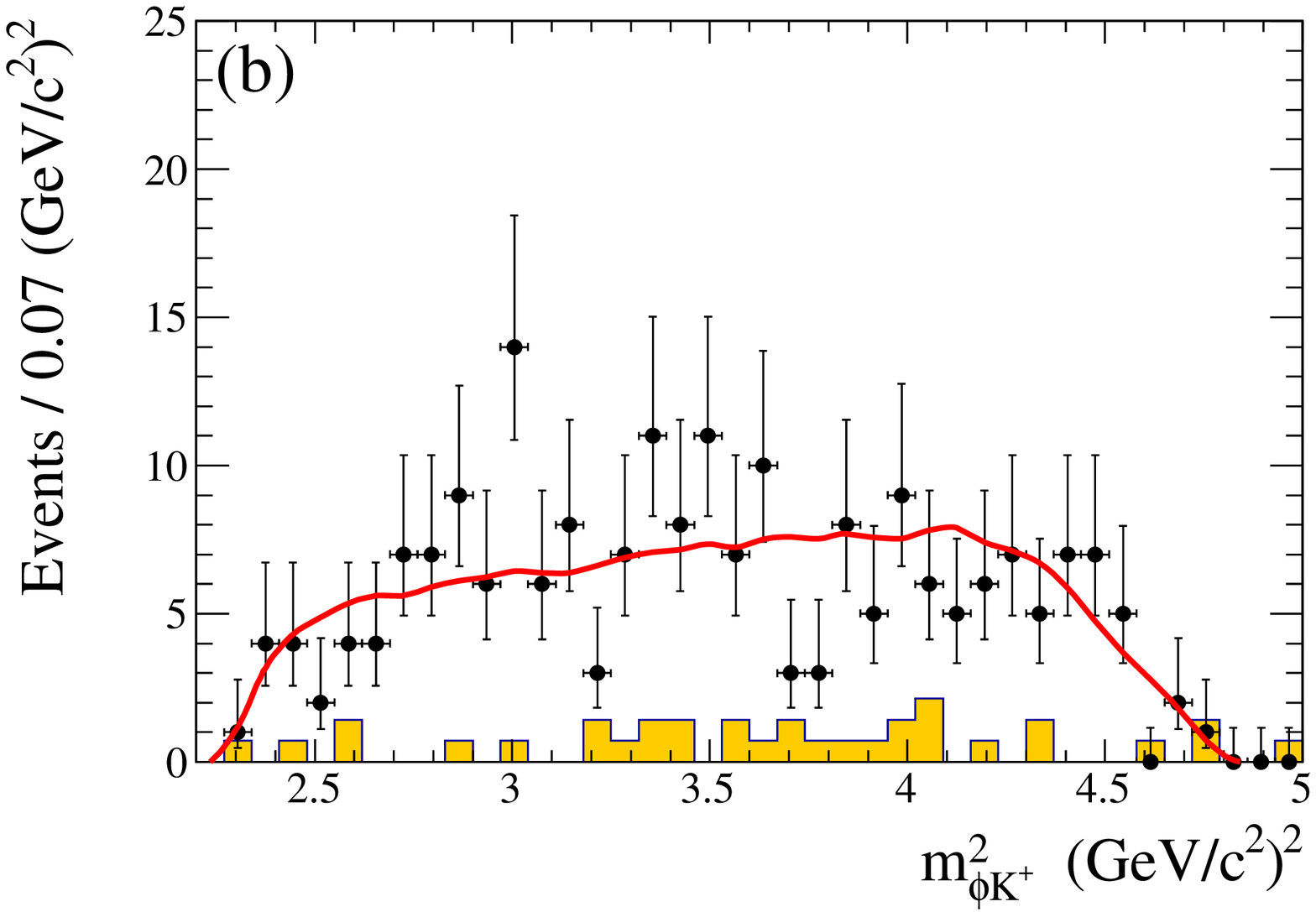}}
\scalebox{0.28}{\includegraphics{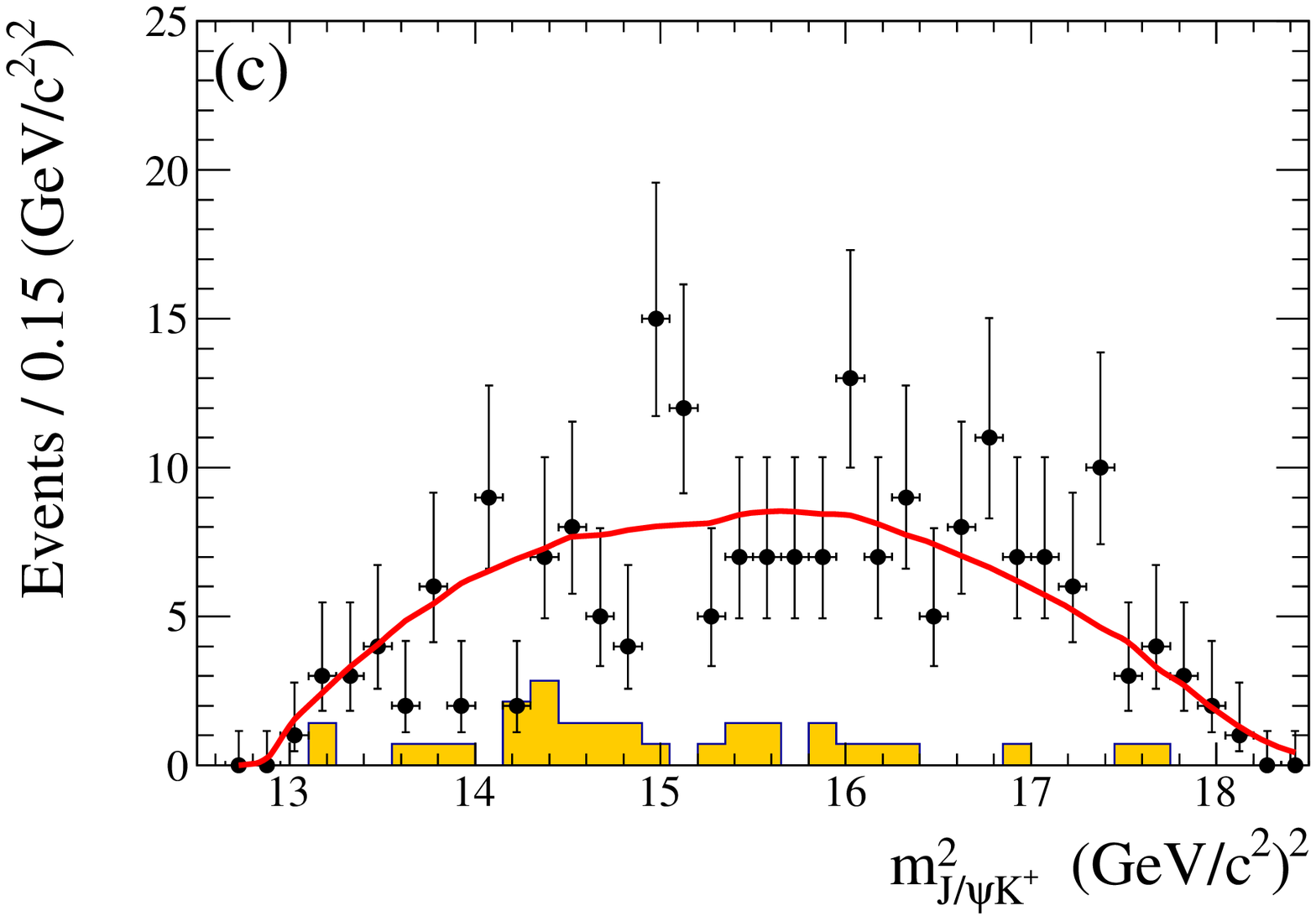}}
}
\caption{Dalitz plot projections for $B^+ \rightarrow J/\psi \phi K^+$ on (a) $m^2_{\jpsi \phi}$, (b) $m^2_{\phi K^+}$, and (c) $m^2_{\jpsi K^+}$. The continuous (red) curves are the results from fit model A performed including the $X(4140)$ and $X(4270)$ resonances. The dashed (blue) curve in (a)  indicates the projection for fit model D, with no resonances. The shaded (yellow) histograms indicate the background estimated from the \DeltaE sidebands.}
\label{fig:fig_fit}
\end{center}
\end{figure*}
\begin{figure*}[ht] 
\begin{center}
\mbox{
\scalebox{0.28}{\includegraphics{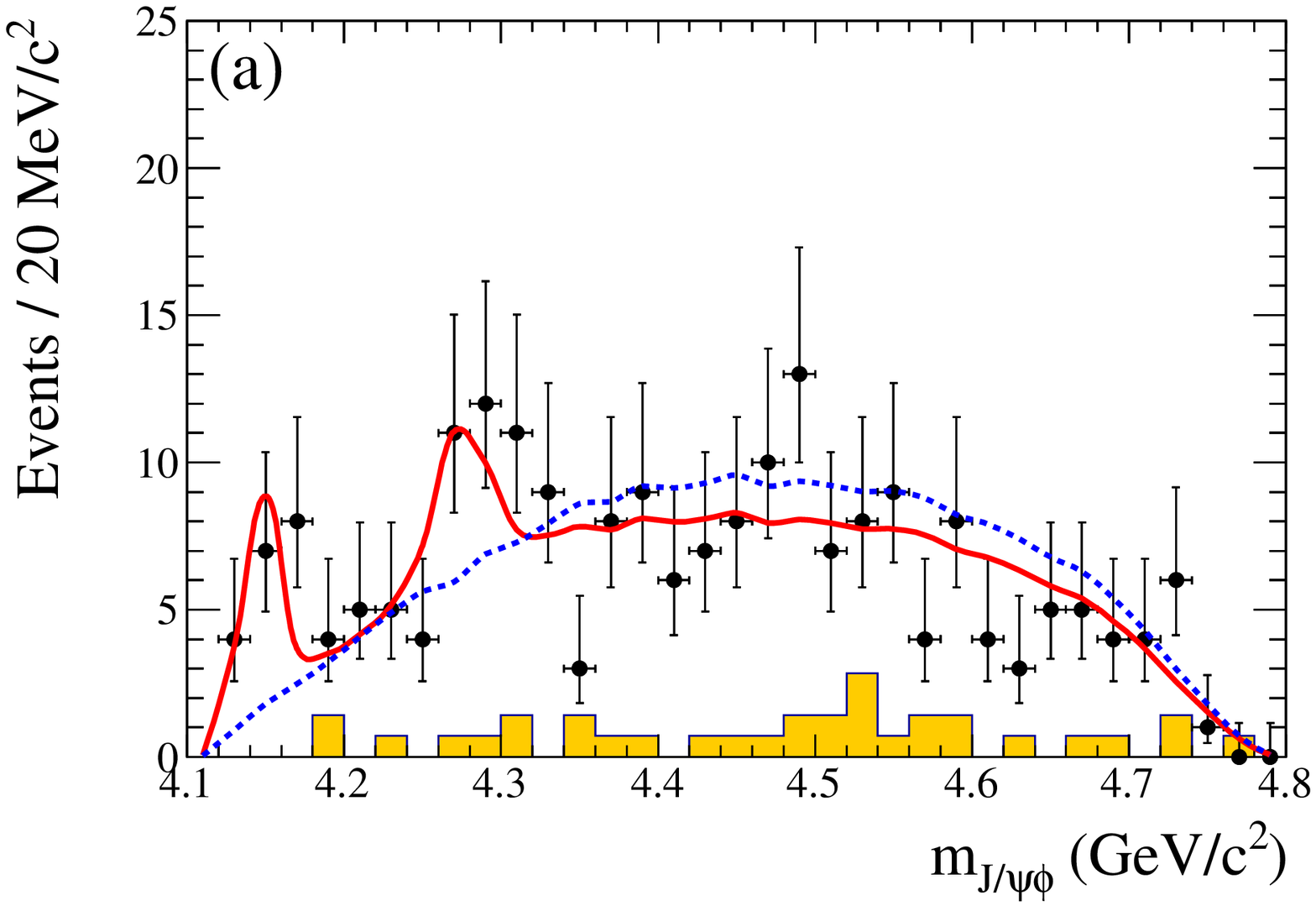}}
\scalebox{0.28}{\includegraphics{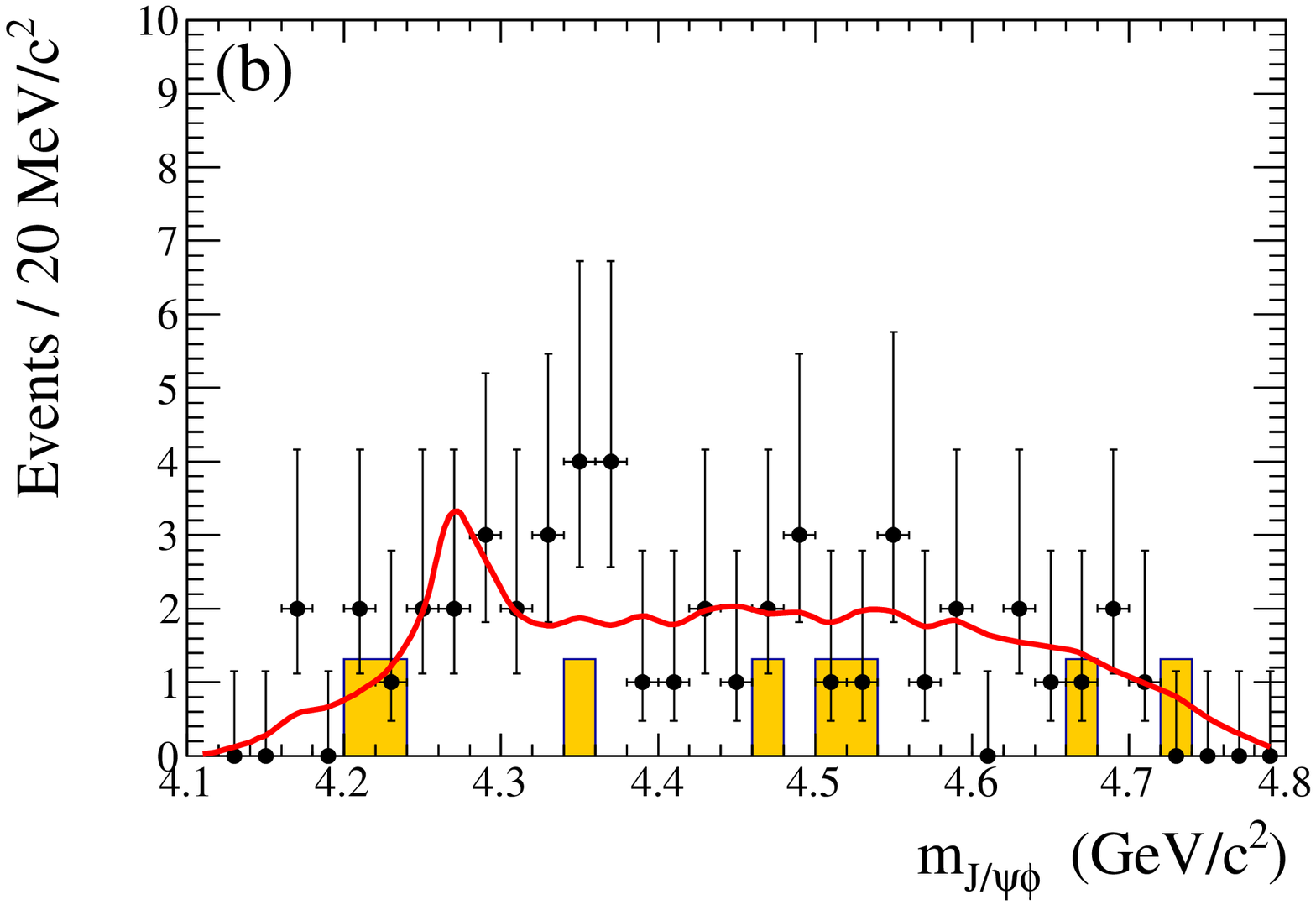}}
\scalebox{0.28}{\includegraphics{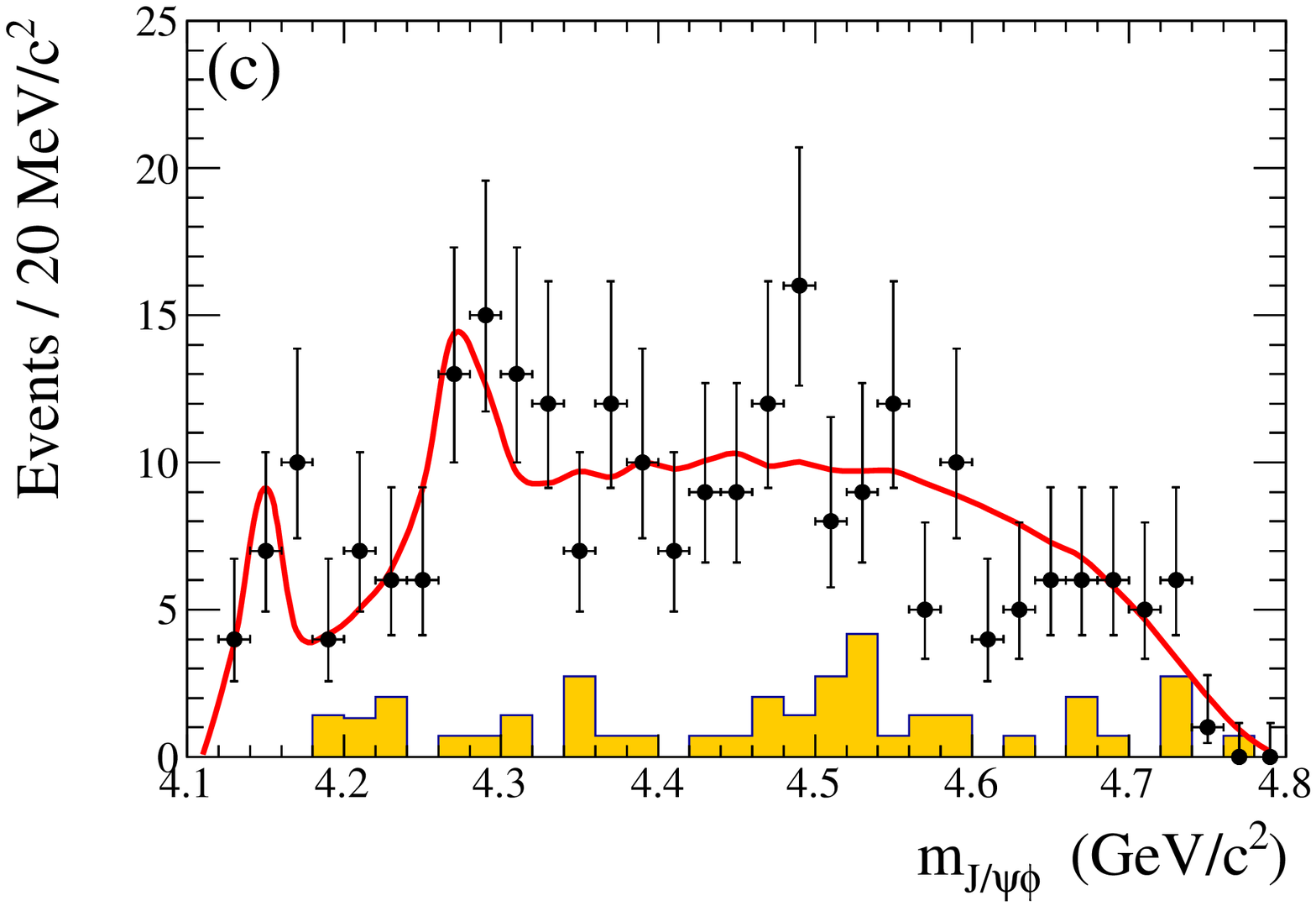}}
}
\caption{Projections on the $J/\psi \phi$ mass spectrum from the Dalitz plot fit with the $X(4140)$ and the $X(4270)$ resonances for the (a) $B^+$, (b) $B^0$, and (c) combined $B^+$ and  $B^0$ data samples. The continuous (red) curves result from the fit; the dashed (blue) curve in (a)  indicates the projection for fit model D, with no resonances. The shaded (yellow) histograms show the background contributions estimated from the \DeltaE sidebands.}
\label{fig:fig_fit_t}
\end{center}
\end{figure*}
\begin{figure*}[ht] 
\begin{center}
\mbox{
\scalebox{0.28}{\includegraphics{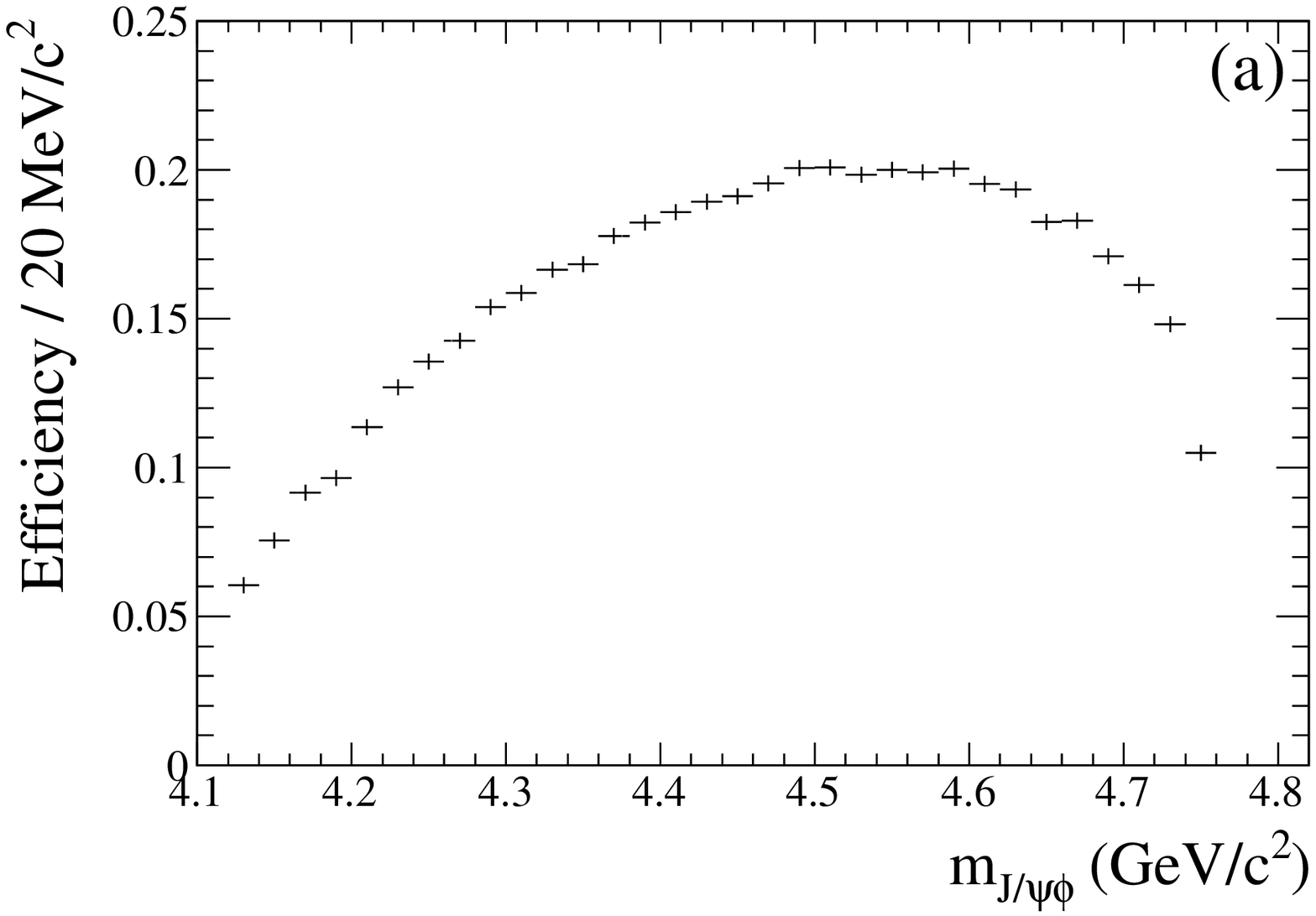}}
\scalebox{0.28}{\includegraphics{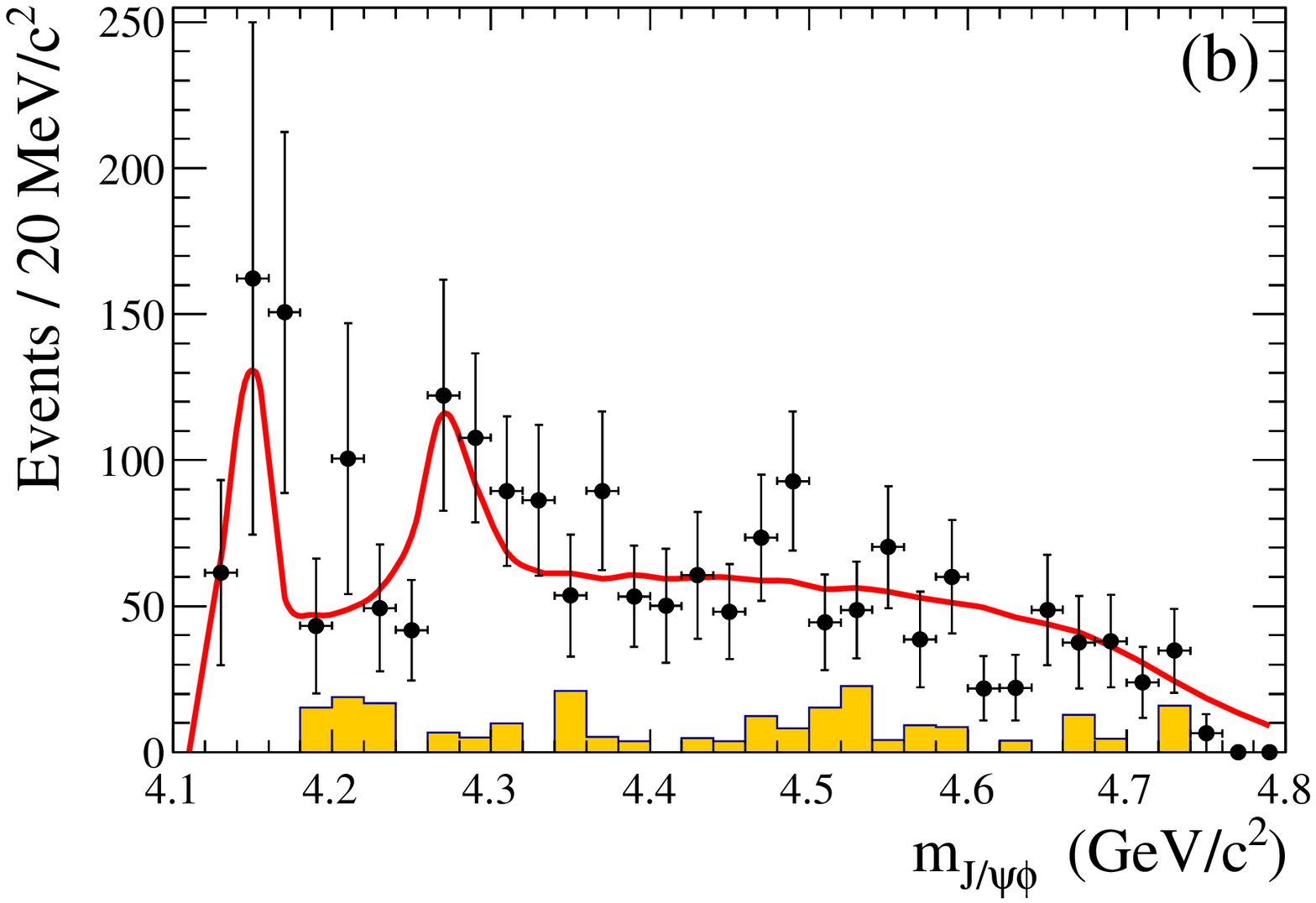}}
\scalebox{0.28}{\includegraphics{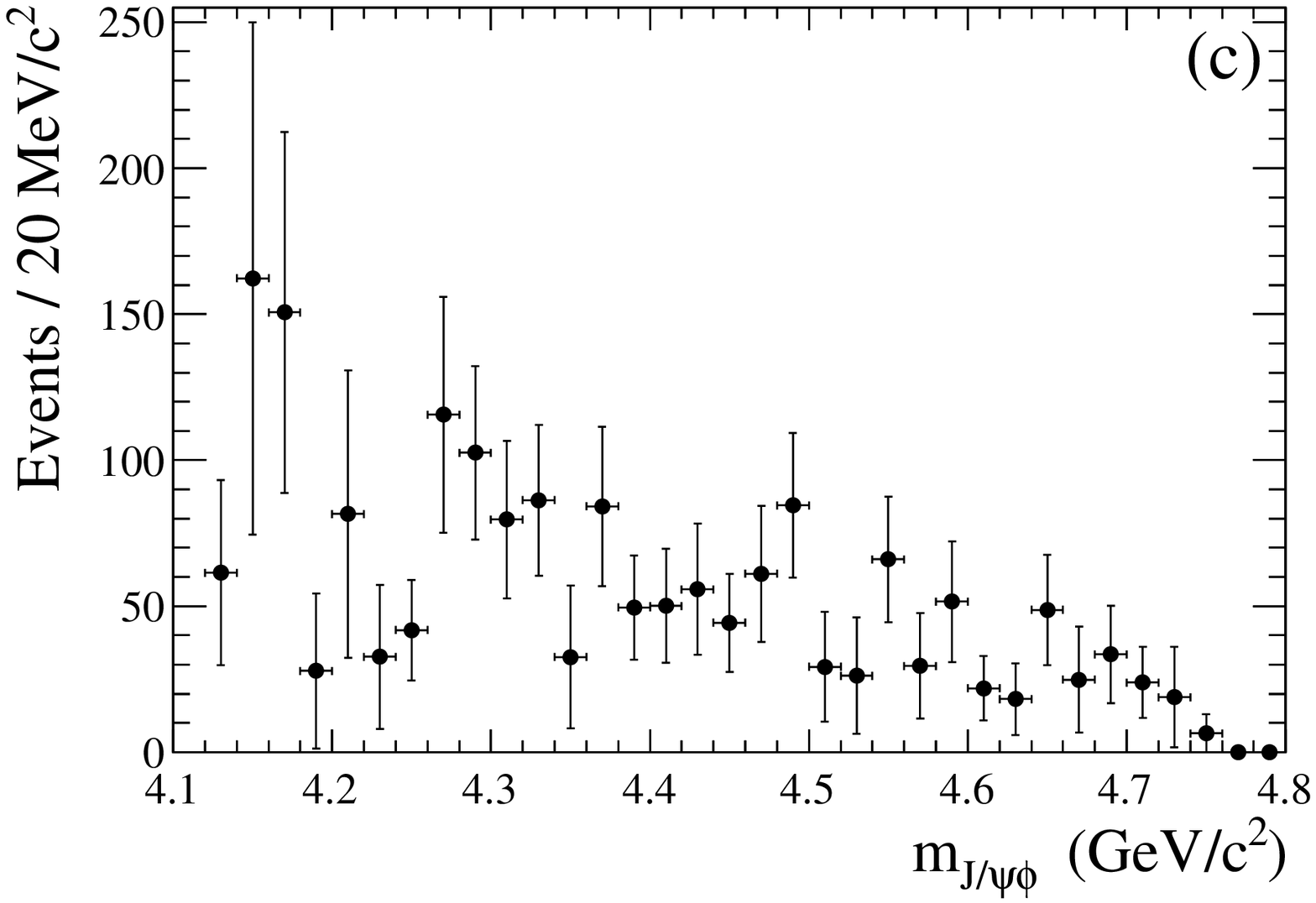}}
}
\caption{(a) Average efficiency distribution as a function of $J/\psi \phi$ invariant mass for $B^+ \rightarrow J/\psi \phi K^+$. (b) Efficiency-corrected $J/\psi \phi$ mass spectrum for the combined $B^+$ and $B^0$ samples. The curve is the result from fit model A described in the text. The shaded (yellow) histogram represents the efficiency-corrected background contribution. (c) Efficiency-corrected and background-subtracted  $J/\psi \phi$ mass spectrum for the combined $B^+$ and $B^0$ samples.}
\label{fig:fig8}
\end{center}
\end{figure*}

We plot in Fig.~\ref{fig4}(a) the $\jpsi \Kp \Km$ mass distribution for  $B^+ \rightarrow J/\psi K^+ K^- K^+$ and in Fig.~\ref{fig4}(b) that for $B^0 \rightarrow J/\psi  K^- K^+ \KS$; the signal regions are defined by the \DeltaE selections indicated in Sec. II and \mes$>5.27$ \gevcc. No prominent structure is observed in both mass spectra.

We select events in the $\phi$ signal regions and search for the resonant states reported by the CDF Collaboration in the
$J/\psi \phi$ mass spectrum~\cite{kai-bis}. The mass and the width values are fixed to
$m$=4143.4 \mevcc and $\Gamma$=15.3 \mev for the $X(4140)$, and $m$=4274.4 \mevcc and $\Gamma$= 32.3 \mev for the $X(4270)$ resonance. 
We evaluate the mass resolution using MC simulations and obtain 2 \mevcc resolution in the mass region between 4100 \mevcc and 4300 \mevcc. Therefore resolution effects can be ignored because they are much smaller than the widths of the resonances under consideration.

We estimate the efficiency on each quasi-three-body Dalitz plot as the ratio between the reconstructed and generated distributions, where the values are generated according to PHSP. Figure~\ref{fig5} shows the resulting distributions evaluated over the $m^2_{\jpsi \phi} ~vs~ m^2_{\phi K}$ plane for the charged (a) and neutral (b) $B$ decay, respectively. The lower efficiency at low $J/\psi \phi$ mass is due to the lower reconstruction efficiency for low kaon momentum in the laboratory frame, as a result of energy loss in the beampipe and SVT material.

We test the agreement between data and MC by using a full MC simulation where the $B^+ \rightarrow J/\psi \phi K^+$ and $B^0 \rightarrow J/\psi \phi \KS$ decays are included with known branching fractions. We repeat the entire analysis
on these simulated data and find good agreement between generated and reconstructed branching fractions.
Resolution effects are small and are computed using MC simulations. We obtain
average values of 2.9 MeV for ($J/\psi \phi$) and 2.2 MeV for ($J/\psi K$). These small values do not produce bias in the evaluation of the efficiency and the measurement of the branching fractions.

To search for the two resonances in the $J/\psi \phi$ mass distributions, we perform unbinned maximum likelihood fits to the $B\rightarrow J/\psi \phi K$ decay Dalitz plots. 
We model the resonances using S-wave relativistic Breit-Wigner (BW) functions with parameters fixed to the CDF values. 
The non-resonant contributions are represented by a constant term, and no interference is allowed between the fit components. 
We estimate the background contributions from the \DeltaE sidebands, find them to be small and consistent with a PHSP behavior,  and so in the fits they are incorporated into the non-resonant PHSP term. The decay of a pseudoscalar meson to two vector states may contain high spin contributions which could generate non-uniform angular distributions. However, due to the limited data sample we do not include such angular terms, and assume that the resonances decay isotropically. The amplitudes are normalized using PHSP MC generated events with $B$ parameters obtained from the fits to the data. The fit functions are weighted by the the two-dimensional efficiency computed on the Dalitz plots. 

We perform fits separately for the charged $B^+$ sample and the combined $B^+$ and \Bz samples. Due to the very limited statistics of the $B^0$ sample we do not perform a separate fit, but instead subtract the fit result for the $B^+$ sample from that for the combined $B^+$ and $B^0$ sample.
In this case we make use of the two different efficiencies for the two channels. In the MC simulation performed, we make use of a weighted mean of the two efficiencies evaluated on the respective Dalitz plots.

Table~III summarizes the results of the fits.
We report the background-corrected fit fractions for the two resonances, $f_{X(4140)}$ and
$f_{X(4270)}$, the two-dimensional (2D) $\chi^2$ computed on the Dalitz plot, and the one-dimensional (1D) $\chi^2$ computed on the $\jpsi \phi$ mass projection.
For this purpose, we use an adaptive binning method, and divide the Dalitz plot into a number of cells in such a way that the minimum expected population per cell is not smaller than 7. We generate MC simulations weighted by the  efficiency and by the results from the fits. These are normalized to the event yield in data, using the same bin definitions.
We then compute the $\chi^2 = \sum_{i=1}^{N_{\rm cells}} (N^i_{\rm obs}-N^i_{\rm exp})^2/N^i_{\rm exp}$ where $N^i_{\rm obs}$ and $N^i_{\rm exp}$ are the data and MC simulation event yields, respectively. Indicating with $n$ the number of free parameters, corresponding to the number of resonances included in the fit, the number of degrees of freedom is $\nu=N_{\rm cells}-n$. In computing the 1D $\chi^2$ we rebin the $\jpsi \phi$ mass projection into 25 bins, again with at least 7 entries per bin.

We perform the fits using models with two resonances (labeled as model A), one resonance (models B and C), and no resonances (model D). The fit projections for fit A are displayed in Fig.~\ref{fig:fig_fit}, showing enhancements with a statistical significance smaller than 3.2$\sigma$ for all fit models. All models provide a reasonably good description of the data, with $\chi^2$ probability larger than 1$\%$.

We estimate systematic uncertainties on the fractions by varying the mass and the width values for both resonances within their uncertainties. The results shown in Table~\ref{tab:tab3} are corrected by the fraction of background estimated in each sample. This results in correction factors of 1.12 and 1.21 for the $B^+$ and the $B^0$ channels, respectively. We obtain the following background-corrected fractions for $B^+$:
\begin{equation}
f_{X(4140)} = (9.2 \pm 3.3 \pm 4.7)\%, \ f_{X(4270)} = (10.6 \pm 4.8 \pm 7.1) \%.
\end{equation}
Combining statistical and systematic uncertainties in quadrature, we obtain significances of 1.6 and 1.2$\sigma$ for the $X(4140)$ and the $X(4270)$, respectively.

Using the Feldman-Cousins method~\cite{FC}, we obtain the ULs at 90\% c.l.:
\begin{eqnarray}
\BR(B^+ \rightarrow X(4140)K^+)\times \BR(X(4140) \rightarrow \jpsi \phi)/ \nonumber\\
\BR(B^+ \rightarrow \jpsi \phi K^+) < 0.133 
\end{eqnarray}
\begin{eqnarray}
\BR(B^+ \rightarrow X(4270)K^+)\times \BR(X(4270)\rightarrow \jpsi \phi)/ \nonumber\\
\BR(B^+ \rightarrow \jpsi \phi K^+) < 0.181.
\end{eqnarray}
The Feldman-Cousin intervals are evaluated as explained in Ref.~\cite{FC} and in Sec. III.
The $X(4140)$ limit may be compared with the CDF measurement of $0.149\pm 0.039\pm 0.024$~\cite{kai} and the LHCb limit of 0.07~\cite{LHCb}. The $X(4270)$ limit may be compared with the LHCb limit of 0.08.

 The fit projections on the $J/\psi \phi$ mass spectrum using  fit model A with two resonances are shown in Fig.~\ref{fig:fig_fit_t}(a) for $B^+$, in Fig.~\ref{fig:fig_fit_t}(b) for $B^0$, and in Fig.~\ref{fig:fig_fit_t}(c) for the combined $B^+$ and $B^0$ sample. The fit results are summarized in Table~\ref{tab:tab3}. 

The central values of mass and width of the two resonances are also fixed to the values recently published by the CMS Collaboration~\cite{cms}. In this case we obtain, for the $B^+$ data, the following background-corrected fractions:

\begin{equation}
f_{X(4140)} = (13.2 \pm 3.8 \pm 6.8)\%, \ f_{X(4270)} = (10.9 \pm 5.2 \pm 7.3) \%. 
\end{equation}
These values are consistent within the uncertainties with those obtained in Eq.~(5).
For comparison, CMS reported a fraction of $0.10 \pm 0.03$ for the X(4140), which is compatible with the CDF, the LHCb and our value within the uncertainties; CMS could not determine reliably the significance of the second structure X(4270) due to possible reflections of two-body decays. 

Figure~\ref{fig:fig8}(a) shows the efficiency as a function of the $J/\psi \phi$ mass, obtained from a PHSP simulation of the $B^+ \to J/\psi \phi K^+$ Dalitz plot. 
We observe a decrease of the efficiency in the $J/\psi \phi$ threshold region, as already observed in Fig.~\ref{fig5}.

Figure~\ref{fig:fig8}(b) shows the efficiency-corrected $J/\psi \phi$ mass spectrum for the combined $B^+$ and $B^0$ samples. To obtain this spectrum, we weight each event
by the inverse of the efficiency evaluated on the respective $B^+$ and $B^0$ Dalitz plots. The curve is the result from fit model A. The background contribution (shown shaded) is estimated from the \DeltaE sidebands, and has also been corrected for efficiency. However, a few background events fall outside the efficiency Dalitz plots, and to these we assign the same efficiency as for $B$ signal events. 

Finally, Fig.~\ref{fig:fig8}(c) shows the efficiency-corrected and background-subtracted $J/\psi \phi$ mass spectrum for the combined $B^+$ and $B^0$ samples.

\section{Summary}

In ~summary, ~we perform ~a study of the decays $B^{+,0} \rightarrow J/\psi K^+ K^- K^{+,0}$ and $B^{+,0} \rightarrow J/\psi \phi K^{+,0}$, and for the latter obtain much-improved BF measurements. For $B^{+} \rightarrow J/\psi K^+ K^- K^{+}$ this is the first measurement. We search for resonance production in the $\jpsi \phi$ mass spectrum and obtain significances below 2$\sigma$ for both the $X(4140)$ and the $X(4270)$ resonances, with systematic uncertainties taken into account. 
Limits on the product Branching Ratio values for these resonances are obtained. We find that the hypothesis that the events are distributed uniformly on the Dalitz plot gives a poorer description of the data.
We also search for $B^0 \rightarrow J/\psi \phi$ and derive an UL on the BF for this decay mode, which is in agreement with theoretical expectations.

\section{Acknowledgements}
We are grateful for the 
extraordinary contributions of our \pep2\ colleagues in
achieving the excellent luminosity and machine conditions
that have made this work possible.
The success of this project also relies critically on the 
expertise and dedication of the computing organizations that 
support \babar.
The collaborating institutions wish to thank 
SLAC for its support and the kind hospitality extended to them. 
This work is supported by the
US Department of Energy
and National Science Foundation, the
Natural Sciences and Engineering Research Council (Canada),
the Commissariat \`a l'Energie Atomique and
Institut National de Physique Nucl\'eaire et de Physique des Particules
(France), the
Bundesministerium f\"ur Bildung und Forschung and
Deutsche Forschungsgemeinschaft
(Germany), the
Istituto Nazionale di Fisica Nucleare (Italy),
the Foundation for Fundamental Research on Matter (The Netherlands),
the Research Council of Norway, the
Ministry of Education and Science of the Russian Federation, 
Ministerio de Ciencia e Innovaci\'on (Spain), and the
Science and Technology Facilities Council (United Kingdom).
Individuals have received support from 
the Marie-Curie IEF program (European Union), the A. P. Sloan Foundation (USA) 
and the Binational Science Foundation (USA-Israel).

\end{document}

%% file: authors_oct2013_bad2479.tex
\author{J.~P.~Lees}
\author{V.~Poireau}
\author{V.~Tisserand}
\affiliation{Laboratoire d'Annecy-le-Vieux de Physique des Particules (LAPP), Universit\'e de Savoie, CNRS/IN2P3,  F-74941 Annecy-Le-Vieux, France}
\author{E.~Grauges}
\affiliation{Universitat de Barcelona, Facultat de Fisica, Departament ECM, E-08028 Barcelona, Spain }
\author{A.~Palano$^{ab}$ }
\affiliation{INFN Sezione di Bari$^{a}$; Dipartimento di Fisica, Universit\`a di Bari$^{b}$, I-70126 Bari, Italy }
\author{G.~Eigen}
\author{B.~Stugu}
\affiliation{University of Bergen, Institute of Physics, N-5007 Bergen, Norway }
\author{D.~N.~Brown}
\author{L.~T.~Kerth}
\author{Yu.~G.~Kolomensky}
\author{M.~J.~Lee}
\author{G.~Lynch}
\affiliation{Lawrence Berkeley National Laboratory and University of California, Berkeley, California 94720, USA }
% ...
\collaboration{The \babar\ Collaboration}
\noaffiliation

%% file: paper_new2.bbl
\begin{thebibliography}{99}
\bibitem{conj}
The use of charge conjugate reactions is implied throughout this work.
\bibitem{formula}
The BFs are calculated by dividing the efficiency-corrected yields in Table~\ref{tab:tab1} by the product of $\rm{ N_{B \bar B} \times \Pi_i BFS_i}$, where $\rm{ BFS_i}$ = BF of particles involved in the relevant $\Upsilon(4S)$ decay and in the corresponding
       $B$ meson decay processes.
\end{thebibliography}
